\documentclass{aastex62}
\usepackage{graphicx}
\usepackage{amssymb}
\usepackage{amsmath}

\usepackage{natbib}
\usepackage{float}
\usepackage{multirow}
\usepackage{blindtext}
\usepackage[textwidth=16mm]{todonotes}
\usepackage{lineno}

\begin{document}

\title{A search for time-dependent astrophysical neutrino emission with IceCube data from 2012 to 2017}

\affiliation{III. Physikalisches Institut, RWTH Aachen University, D-52056 Aachen, Germany}
\affiliation{Department of Physics, University of Adelaide, Adelaide, 5005, Australia}
\affiliation{Dept. of Physics and Astronomy, University of Alaska Anchorage, 3211 Providence Dr., Anchorage, AK 99508, USA}
\affiliation{Dept. of Physics, University of Texas at Arlington, 502 Yates St., Science Hall Rm 108, Box 19059, Arlington, TX 76019, USA}
\affiliation{CTSPS, Clark-Atlanta University, Atlanta, GA 30314, USA}
\affiliation{School of Physics and Center for Relativistic Astrophysics, Georgia Institute of Technology, Atlanta, GA 30332, USA}
\affiliation{Dept. of Physics, Southern University, Baton Rouge, LA 70813, USA}
\affiliation{Dept. of Physics, University of California, Berkeley, CA 94720, USA}
\affiliation{Lawrence Berkeley National Laboratory, Berkeley, CA 94720, USA}
\affiliation{Institut f{\"u}r Physik, Humboldt-Universit{\"a}t zu Berlin, D-12489 Berlin, Germany}
\affiliation{Fakult{\"a}t f{\"u}r Physik {\&} Astronomie, Ruhr-Universit{\"a}t Bochum, D-44780 Bochum, Germany}
\affiliation{Universit{\'e} Libre de Bruxelles, Science Faculty CP230, B-1050 Brussels, Belgium}
\affiliation{Vrije Universiteit Brussel (VUB), Dienst ELEM, B-1050 Brussels, Belgium}
\affiliation{Department of Physics and Laboratory for Particle Physics and Cosmology, Harvard University, Cambridge, MA 02138, USA}
\affiliation{Dept. of Physics, Massachusetts Institute of Technology, Cambridge, MA 02139, USA}
\affiliation{Dept. of Physics and Institute for Global Prominent Research, Chiba University, Chiba 263-8522, Japan}
\affiliation{Department of Physics, Loyola University Chicago, Chicago, IL 60660, USA}
\affiliation{Dept. of Physics and Astronomy, University of Canterbury, Private Bag 4800, Christchurch, New Zealand}
\affiliation{Dept. of Physics, University of Maryland, College Park, MD 20742, USA}
\affiliation{Dept. of Astronomy, Ohio State University, Columbus, OH 43210, USA}
\affiliation{Dept. of Physics and Center for Cosmology and Astro-Particle Physics, Ohio State University, Columbus, OH 43210, USA}
\affiliation{Niels Bohr Institute, University of Copenhagen, DK-2100 Copenhagen, Denmark}
\affiliation{Dept. of Physics, TU Dortmund University, D-44221 Dortmund, Germany}
\affiliation{Dept. of Physics and Astronomy, Michigan State University, East Lansing, MI 48824, USA}
\affiliation{Dept. of Physics, University of Alberta, Edmonton, Alberta, Canada T6G 2E1}
\affiliation{Erlangen Centre for Astroparticle Physics, Friedrich-Alexander-Universit{\"a}t Erlangen-N{\"u}rnberg, D-91058 Erlangen, Germany}
\affiliation{Physik-department, Technische Universit{\"a}t M{\"u}nchen, D-85748 Garching, Germany}
\affiliation{D{\'e}partement de physique nucl{\'e}aire et corpusculaire, Universit{\'e} de Gen{\`e}ve, CH-1211 Gen{\`e}ve, Switzerland}
\affiliation{Dept. of Physics and Astronomy, University of Gent, B-9000 Gent, Belgium}
\affiliation{Dept. of Physics and Astronomy, University of California, Irvine, CA 92697, USA}
\affiliation{Karlsruhe Institute of Technology, Institute for Astroparticle Physics, D-76021 Karlsruhe, Germany }
\affiliation{Dept. of Physics and Astronomy, University of Kansas, Lawrence, KS 66045, USA}
\affiliation{SNOLAB, 1039 Regional Road 24, Creighton Mine 9, Lively, ON, Canada P3Y 1N2}
\affiliation{Department of Physics and Astronomy, UCLA, Los Angeles, CA 90095, USA}
\affiliation{Department of Physics, Mercer University, Macon, GA 31207-0001, USA}
\affiliation{Dept. of Astronomy, University of Wisconsin{\textendash}Madison, Madison, WI 53706, USA}
\affiliation{Dept. of Physics and Wisconsin IceCube Particle Astrophysics Center, University of Wisconsin{\textendash}Madison, Madison, WI 53706, USA}
\affiliation{Institute of Physics, University of Mainz, Staudinger Weg 7, D-55099 Mainz, Germany}
\affiliation{Department of Physics, Marquette University, Milwaukee, WI, 53201, USA}
\affiliation{Institut f{\"u}r Kernphysik, Westf{\"a}lische Wilhelms-Universit{\"a}t M{\"u}nster, D-48149 M{\"u}nster, Germany}
\affiliation{Bartol Research Institute and Dept. of Physics and Astronomy, University of Delaware, Newark, DE 19716, USA}
\affiliation{Dept. of Physics, Yale University, New Haven, CT 06520, USA}
\affiliation{Dept. of Physics, University of Oxford, Parks Road, Oxford OX1 3PU, UK}
\affiliation{Dept. of Physics, Drexel University, 3141 Chestnut Street, Philadelphia, PA 19104, USA}
\affiliation{Physics Department, South Dakota School of Mines and Technology, Rapid City, SD 57701, USA}
\affiliation{Dept. of Physics, University of Wisconsin, River Falls, WI 54022, USA}
\affiliation{Dept. of Physics and Astronomy, University of Rochester, Rochester, NY 14627, USA}
\affiliation{Oskar Klein Centre and Dept. of Physics, Stockholm University, SE-10691 Stockholm, Sweden}
\affiliation{Dept. of Physics and Astronomy, Stony Brook University, Stony Brook, NY 11794-3800, USA}
\affiliation{Dept. of Physics, Sungkyunkwan University, Suwon 16419, Korea}
\affiliation{Institute of Basic Science, Sungkyunkwan University, Suwon 16419, Korea}
\affiliation{Dept. of Physics and Astronomy, University of Alabama, Tuscaloosa, AL 35487, USA}
\affiliation{Dept. of Astronomy and Astrophysics, Pennsylvania State University, University Park, PA 16802, USA}
\affiliation{Dept. of Physics, Pennsylvania State University, University Park, PA 16802, USA}
\affiliation{Dept. of Physics and Astronomy, Uppsala University, Box 516, S-75120 Uppsala, Sweden}
\affiliation{Dept. of Physics, University of Wuppertal, D-42119 Wuppertal, Germany}
\affiliation{DESY, D-15738 Zeuthen, Germany}

\author{R. Abbasi}
\affiliation{Department of Physics, Loyola University Chicago, Chicago, IL 60660, USA}
\author{M. Ackermann}
\affiliation{DESY, D-15738 Zeuthen, Germany}
\author{J. Adams}
\affiliation{Dept. of Physics and Astronomy, University of Canterbury, Private Bag 4800, Christchurch, New Zealand}
\author{J. A. Aguilar}
\affiliation{Universit{\'e} Libre de Bruxelles, Science Faculty CP230, B-1050 Brussels, Belgium}
\author{M. Ahlers}
\affiliation{Niels Bohr Institute, University of Copenhagen, DK-2100 Copenhagen, Denmark}
\author{M. Ahrens}
\affiliation{Oskar Klein Centre and Dept. of Physics, Stockholm University, SE-10691 Stockholm, Sweden}
\author{C. Alispach}
\affiliation{D{\'e}partement de physique nucl{\'e}aire et corpusculaire, Universit{\'e} de Gen{\`e}ve, CH-1211 Gen{\`e}ve, Switzerland}
\author{A. A. Alves Jr.}
\affiliation{Karlsruhe Institute of Technology, Institute for Astroparticle Physics, D-76021 Karlsruhe, Germany }
\author{N. M. Amin}
\affiliation{Bartol Research Institute and Dept. of Physics and Astronomy, University of Delaware, Newark, DE 19716, USA}
\author{K. Andeen}
\affiliation{Department of Physics, Marquette University, Milwaukee, WI, 53201, USA}
% ...
\date{\today}

\collaboration{IceCube Collaboration}
\noaffiliation

\begin{abstract}

High-energy neutrinos are unique messengers of the high-energy universe, tracing the processes of cosmic-ray acceleration. This paper presents analyses focusing on time-dependent neutrino point-source searches. A scan of the whole sky, making no prior assumption about source candidates, is performed, looking for a space and time clustering of high-energy neutrinos in data collected by the IceCube Neutrino Observatory between 2012 and 2017. No statistically significant evidence for a time-dependent neutrino signal is found with this search during this period since all results are consistent with the background expectation. 
Within this study period, the blazar 3C 279, showed strong variability, inducing a very prominent gamma-ray flare observed in 2015 June. This event motivated a dedicated study of the blazar, which consists of searching for a time-dependent neutrino signal correlated with the gamma- ray emission. No evidence for a time-dependent signal is found. Hence, an upper limit on the neutrino fluence is derived, allowing us to constrain a hadronic emission model.

\end{abstract}



\section{Introduction} \label{sec:intro}

The cosmic-ray energy spectrum is observed from the sub-GeV region to ultra high energies in the $10^{20}$~eV range. While the lowest-energy cosmic rays are thought to originate from Galactic supernova explosions \citep{PhysRev.46.76.2}, extragalactic objects, such as gamma-ray bursts and active galactic nuclei \citep{Waxman:1998yy,Mannheim:1998wp}, are candidates to explain the highest energy part of the spectrum. However, none of these hypotheses has been experimentally verified and the origin of cosmic rays remains mostly unknown.

Cosmic rays themselves don't point back at their exact origin because they are deflected by magnetic fields on their way to Earth. It is however possible to get this information through some of the secondary particles cosmic rays produce in the vicinity of their sources.  Indeed, cosmic rays interact with gas and radiation near their accelerating sites, leading to the production of charged and neutral pions, which in turn produce neutrinos and photons. Those two particles are indirect messengers that have the potential to point back at cosmic ray sources, as they are both neutral and thus not sensitive to magnetic fields.
However, photon detection alone is delicate; many different phenomena can be responsible for their emission and the photons detected at Earth or by satellites are not necessarily the same as the ones that were produced by cosmic-ray interaction. Photons are also produced in leptonic processes such as bremsstrahlung, inverse Compton scattering and synchrotron radiation, which would be overlaid to the photons produced in hadronic processes, alongside the neutrinos. 
Neutrinos, on the other hand, can only have hadronic origin. They are powerful messengers since they pass through the universe basically undisturbed. At the same time, this implies a challenge for their detection.
The observation of both neutrinos and gamma rays from a same source is a smoking gun to trace cosmic-ray sources.

The IceCube detector has been able to address the neutrino challenge and has reported the detection of a cosmic neutrino flux at the level of few events per year above $\sim 50$~TeV on top of the softer atmospheric neutrino and muon background~\citep{Aartsen:2013jdh,Aartsen:2013eka,aartsen2015atmospheric}. 
Even if the sources of those high-energy neutrinos are still unknown, some evidence for possible candidates started to show up in recent observations. In 2017, a very high-energy neutrino event, IceCube-170922A, was detected. It triggered an alert and a follow-up in the event's direction and in various photon bands was realized by several independent experiments. Evidence for a blazar in a flaring state, TXS 0506+056, consistent with the direction of the neutrino event, was first claimed by the \emph{Fermi} gamma-ray satellite, then confirmed by the MAGIC telescopes and also other high-energy gamma-ray experiments \citep{IceCube:2018dnn, Abeysekara_2018}. The analysis reported a coincidence between  IceCube-170922A and the flare of TXS 0506+056 which was inconsistent with a background fluctuation at the level of 3$\sigma$. 
In addition to that, a time-dependent search, similar to what is described in this paper, was conducted on 9.5 years of archival data and revealed a 3.5$\sigma$ evidence for neutrino emission in the direction of this blazar during a $\sim110$-day period in 2014-2015 \citep{IceCube:2018cha}, this time not accompanied by an observed blazar flare. Those observations combined together make TXS 0506+056 the first blazar associated with a significant neutrino excess. However, only $\sim 1\%$ of the diffuse flux previously observed in IceCube can be attributed to this single source~\citep{2018arXiv181107979I}. 

In 2019, a search for astrophysical neutrino point-like sources was conducted on 10 years of IceCube data. Several types of analysis were realized. Very interestingly, the scan of the whole sky, using neutrino-data only, revealed a clustering of high-energy neutrinos, exceeding the background expectation at the level of 2.9$\sigma$, at a location in the sky coinciding with the Seyfert II galaxy NGC 1068 \citep{Aartsen:2019fau}. In addition to the all-sky analysis, a catalog of sources observed in gamma rays was constructed. All sources considered together, the catalog of the Northern hemisphere was found to be inconsistent with a background-only hypothesis at 3.3$\sigma$. This result is mostly due to excesses found in the neutrino data in the directions of the Seyfert II galaxy NGC 1068, the blazar TXS 0506+056, and the BL Lacs PKS 1424+240 and GB6 J1542+6129. Those results, together with flaring observations from the direction of TXS 0506+056, motivate searches for neutrino sources whose emission is time-dependent, which can provide complementary information to the time-integrated analyses.

In this paper we illustrate a dedicated search for time-dependent signals performed on IceCube data from 2012 to 2017, which consists of looking for a space and time clustering of high-energy neutrinos above 1 TeV, scanning the whole sky. It updates the previous search which used IceCube data from 2008 to 2012 \citep{Aartsen:2015wto}.
Time-dependent studies are relevant since, for flares below about 100 days, they have a better discovery potential than time-integrated searches \citep{Aartsen:2015wto}.  

One time-variable source of special interest is the blazar 3C 279, a frequently studied flat spectrum radio quasar (FSRQ) at redshift $z = 0.536$. It has been observed since the end of the 1980s in radio \citep{aller1987evidence}, optical and infrared \citep{1990Ap&SS.171...13K} and X-ray bands \citep{1989ApJ...347L...9M} and the first gamma-ray observation was achieved in 1991 by the EGRET experiment \citep{hartman1992detection,maraschi1992jet}. More recently, the \textit{Fermi}-Large Area Telescope (LAT) detected two prominent gamma-ray flares, in December 2013 \citep{2015ApJ...807...79H} and on 2015 June 16, the latter being exceptionally bright \citep{2015ATel76331C,2015ApJ...807...79H}, thus making this blazar one of the brightest objects ever observed at such high energies and such distances. For one day, the \textit{Fermi}-LAT experiment detected a flux exceeding the steady flux of the source by more than forty times. This is the historically highest gamma-ray luminosity observed from 3C 279, including also past EGRET observations, with a $\gamma$ ray isotropic luminosity reaching $\sim 10^{49}$ erg~s$^{-1}$. Moreover, a minute-timescale variability of the blazar is reported \citep{ackermann2016minute}, with the flux doubling in less than 10 minutes, which is a very rare event. This allows us to explore compatibility of observations with purely leptonic standard models of FSRQs, in which gamma rays are produced by external radiation comptonization (ERC) \citep{Sikora_2009} or hadronic models. In the ERC model, depending on assumptions, a high Lorentz factor ($>50$) and extremely low jet magnetization or a Lorentz factor of 120 are required \citep{ackermann2016minute}. An alternative leptonic model would imply a magnetically dominated jet still with a high Lorentz factor of about 25 \citep{ackermann2016minute}. 
In this paper, we focus on hadronic models, which, as explained above, foresee interactions of protons with ordinary matter or radiation. In particular, we constrain the model in \citet{halzen2016high}. The analysis of IceCube data, driven by the \textit{Fermi}-LAT gamma-ray data, gives an opportunity to test this model for the 2015 flare of 3C 279. 

\section{IceCube detector and data sample}
\label{sec:ICdetectorDataSets}

The IceCube detector is a cubic kilometer volume of instrumented ice located at the geographic South Pole. Its construction was completed in the austral summer  2010-2011. The detector consists of 86 strings, each holding 50 digital optical modules (DOMs) deployed at a depth between 1450~m and 2450~m below the ice surface. The DOMs are spaced vertically by 17~m and each one hosts a 10-inch photomultiplier tube \citep{Abbasi:2010vc} and associated electronics \citep{Abbasi:2008aa} protected from the high pressure by a glass sphere. The horizontal separation between strings is 125~m. The detection is indirect, i.e. neutrinos interact with the Antarctic ice and produce secondary charged particles such as muons or electrons. Photomultiplier tubes (PMTs) in the DOMs observe Cherenkov radiation from those charged particles. The PMT signals are digitized in the DOMs, and transmitted to the surface.

Muon neutrinos can undergo charged-current interactions in the vicinity or inside the instrumented region, resulting in muon tracks. Cascades originate from charged-current interactions of $\nu_e$ and $\nu_\tau$ and from neutral-current interactions of all neutrino flavours. Track-like events have an angular resolution typically $\leq 0.8^\circ{°}$ at $\sim$ TeV energies and poor energy resolution of the order of a factor of two, while cascade events have about 15\% deposited energy resolution, but only 10$^\circ{°}$ angular resolution above some tens of TeV \citep{energyRecoMethods2014}.
The analyses reported in this paper rely on good pointing capabilities and therefore uses a sample of muon-track events. Neutrino event candidates arriving from the Northern hemisphere (up-going events) are mostly composed of mis-reconstructed down-going muon events and of atmospheric neutrinos with energies above about 100~GeV. The latter constitute the dominant background in this hemisphere. The Southern hemisphere is dominated by down-going atmospheric muons. Therefore a higher energy threshold is applied to reduce them \citep{aartsen2017all}.

The event selections used in the time-dependent analysis are different for the first 3 years and the last 2.5 years of data. Therefore, the two data sets are treated independently and the analysis is performed separately on each of them. Both data sets were recorded with the full configuration of IceCube.
The all-sky scan uses a sample of about $3.5 \times 10^{5}$ good quality tracks (``through-going tracks'') in the Southern and the Northern hemispheres collected from 2012 May to 2015 May. For a detailed description of the event sample, see \citet{aartsen2014searches}. To this sample, also the Medium Energy Starting Events (MESE) lower-energy sample is added \citep{Aartsen_2016}.
The MESE down-going sample is selected by eliminating tracks, mostly due to atmospheric muons, starting outside a fiducial volume inside the instrumented region. In this way, it is possible to select neutrino events with their vertex in the fiducial region. 

From 2015 April to 2017 November a slightly different selection of muon tracks is used: the real-time event selection developed for the gamma-ray follow-up (GFU) program \citep{aartsen2017icecube}. This sample includes $5.7\times 10^{5}$ tracks.

For the analysis of the blazar 3C 279, we use a subsample of 11 days of the through-going tracks and the MESE sample, centered around the bright flare of the blazar, which happened on 2015 June 16. Given that the data selection is the same, the background probability functions (see Sec.~\ref{sec:3C279FlareAnaMethod}) for the blazar analysis were built based on the full period of data between 2012 and 2015. All samples are summarized in Tab.~\ref{tab:icecube_data}.
 
{\centering
\begin{table}
\caption{Summary of the data used in the reported analyses. Each line corresponds to one of the analysis, with the columns giving the specific data selection used, the number of tracks included, the start and end dates of the data and the livetime of the data.} \label{tab:icecube_data}
\begin{tabular}{ccccccc}
\hline \hline
\multirow{2}{*}{Analysis} & Number through-going & Number MESE & Number GFU & Start day & End day & Lifetime \\
 & tracks & tracks & tracks & & & (days)\\
\hline
All-sky scan I &  338588 & 603 & - & 15.05.2012 & 18.05.2015 & 1057.54 \\
All-sky scan II & - & - & 571595 & 24.04.2015 & 01.11.2017 & 886.90\\
3C 279 flare analysis & 3429 & 6 & - & 11.06.2015 & 22.06.2015 & 12.04\\
\hline
\end{tabular}
\end{table}
}

\begin{figure}[h]
\begin{center}
\includegraphics[scale=0.3]{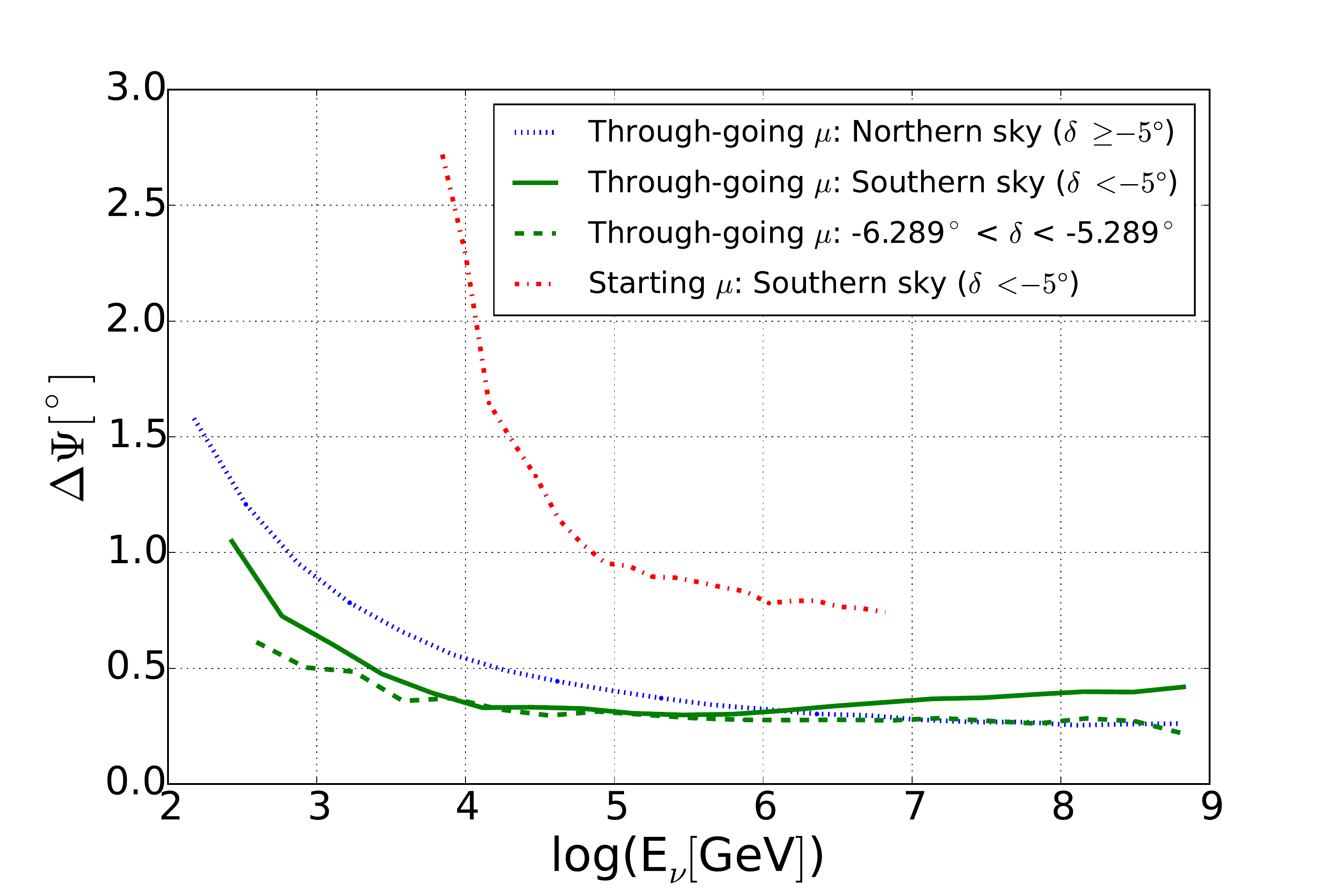}
\end{center}
\caption{Median angular resolution (defined as the median angle between the reconstructed muon track and the parent neutrino direction from simulation) as a function of neutrino energy for the through-going muon tracks data set shown for Northern declinations (dashed blue line), for Southern declinations (solid green line), for a declination band of 1$^\circ{}$ around 3C 279 declination (dashed green line) and for the starting muon tracks (MESE) data set (dashed red line).}
\label{ang_resolution}
\end{figure}

The analysis performance strongly depends on the pointing precision and the effective area, which are described in \citet{Abbasi_2011}.
The pointing accuracy is estimated through the median of the distribution of the angle between the true neutrino direction and the reconstructed muon obtained from Monte Carlo simulations as a function of energy.
Fig.~\ref{ang_resolution} shows this angle for through-going tracks in both the Southern (green) and the Northern (dashed blue) sky, as well as for the declination band of 1$^\circ{°}$ around the blazar 3C 279 (dashed green line). Above 10 TeV, the resolution is below 0.5$^\circ{°}$. The MESE angular resolution is shown in red. It is worse than for the through-going tracks due to the fact that events starting inside the detector have a shorter lever arm. 

\begin{figure}[H]
\begin{center}
\begin{minipage}{0.49\linewidth}

\includegraphics[width=1\textwidth]{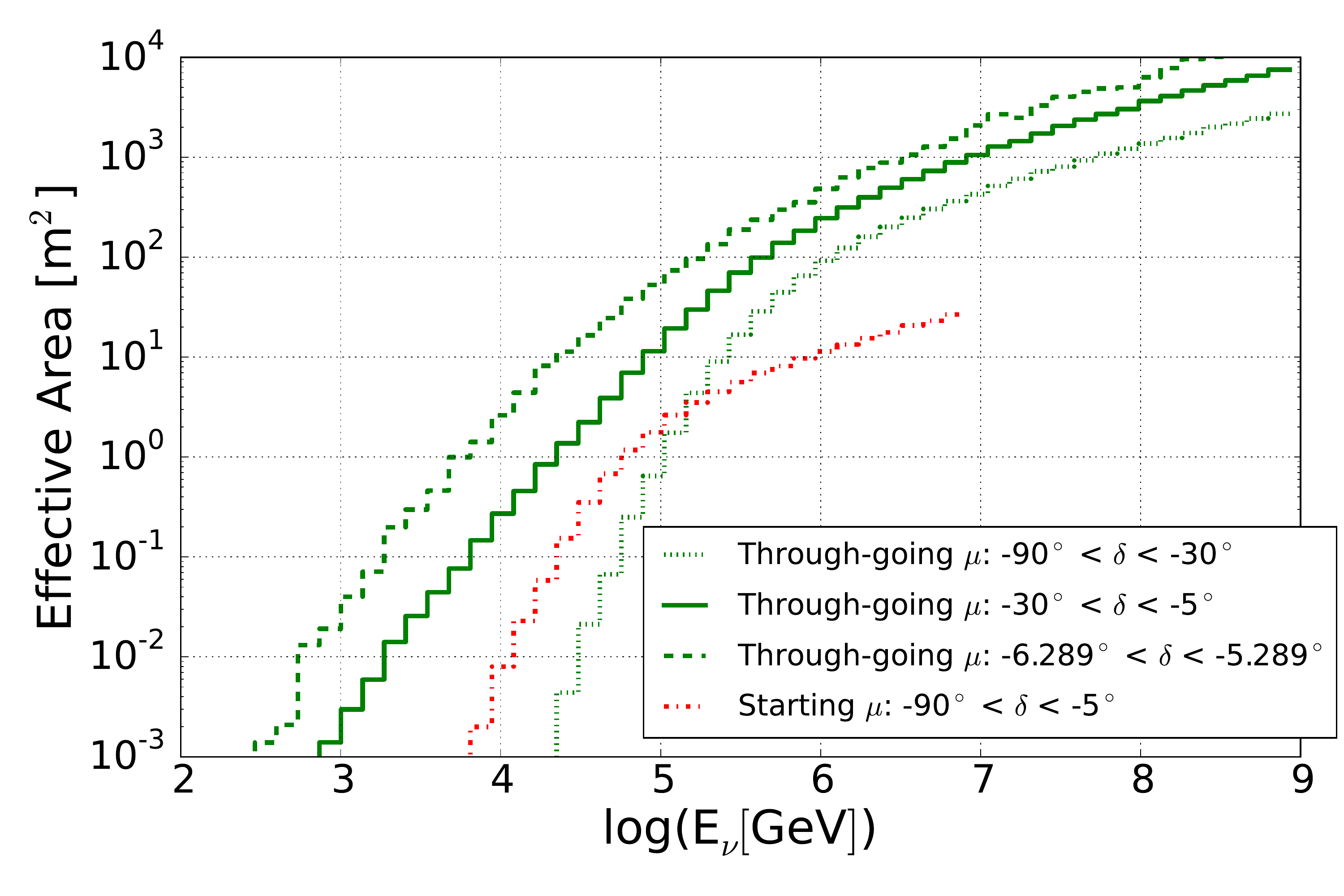}\\ 
\end{minipage}
\hfill
\begin{minipage}{0.49\linewidth}

\includegraphics[width=1.1\textwidth]{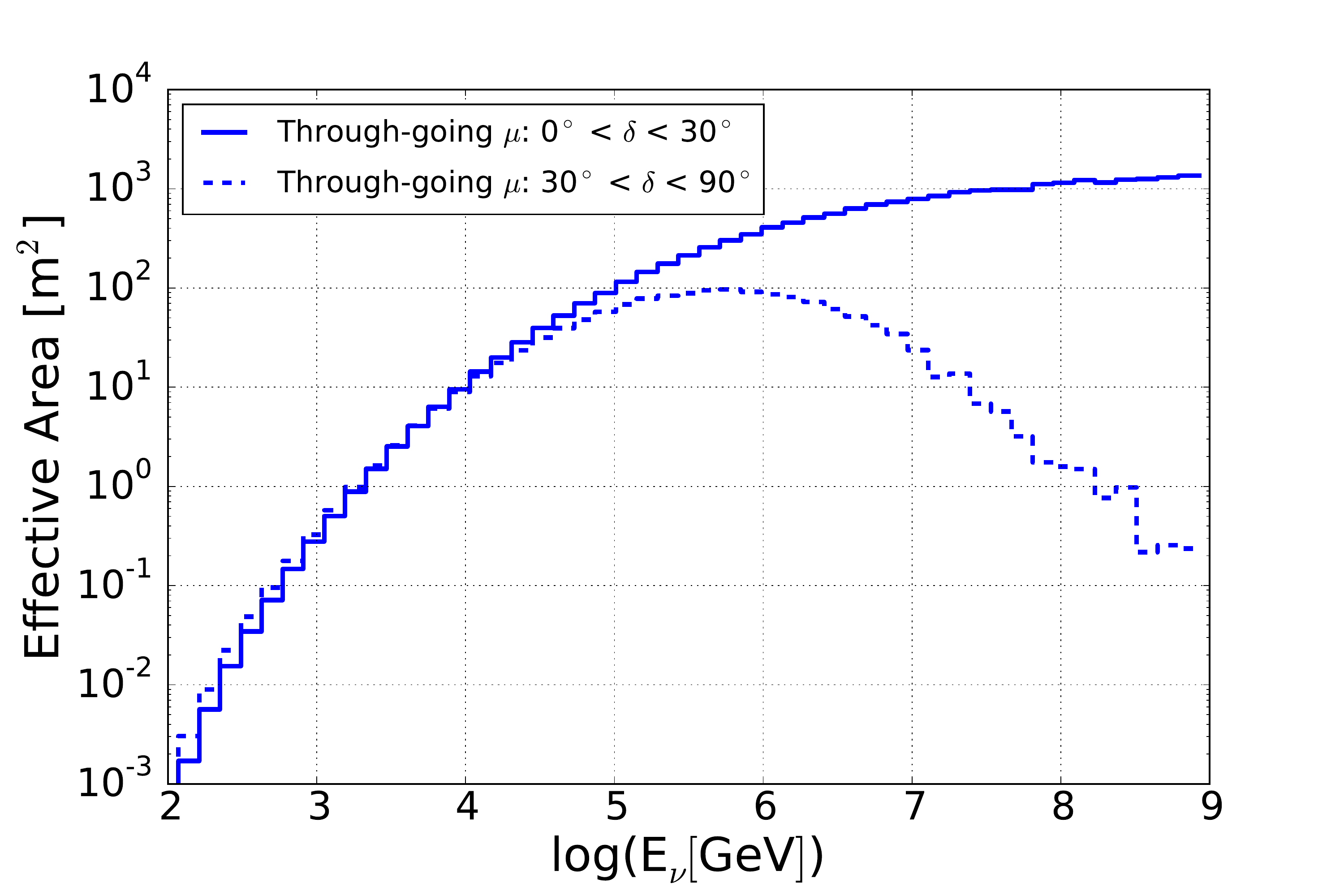}\\ 
\end{minipage}
\caption{IceCube effective area as a function of neutrino energy for the through-going muon and starting muon (MESE) tracks data sets in the Southern hemisphere (left) and for the through-going muon tracks data set in the Northern hemisphere (right). The effective area is averaged over the solid angle in each declination range.}
\label{fig:effective_area}
\end{center}
\end{figure}

The angular resolution of the GFU data set is very similar to the through-going tracks and is shown in \citet{aartsen2017icecube}. The direction of the tracks is calculated through an iterative maximum likelihood method. The uncertainty of the reconstructed direction is estimated by the 1$\sigma$ contour of the likelihood space around the minimum, which is estimated by a method fitting a two-dimensional parabola, called ``paraboloid'', to the likelihood landscape \citep{neunhoffer:2004ha}. Fig.~\ref{fig:effective_area} shows the effective area for the through-going and the MESE selections in the Southern hemisphere (left) and in the Northern hemisphere (right). The effective area is optimal at the location of 3C 279 (green line), since the source is very close to the horizon where the Earth absorption of high-energy neutrinos is negligible. The effective area for the GFU sample is comparable to that of the through-going tracks, as shown in Fig. 3 in~\citet{aartsen2017icecube}.

\section{Analysis method} \label{sec:analysis_method}

The analysis is based on a standard point-source unbinned likelihood method which compares two hypotheses \citep{braun2008methods,braun2010time,Aartsen:2015wto}:
\begin{itemize}
 \item $\mathcal{H}_{0}$, the null hypothesis, which assumes that all events in the data set are due to an isotropic background. 
 
 \item $\mathcal{H}_{1}$, which assumes that measured data are the result of the above described background and in addition a point-source astrophysical neutrino signal following a $\frac{dN}{dE} \propto E^{-\gamma}$ spectrum.

\end{itemize}

To quantify the probability that the data is not compatible with the null hypothesis, a test statistic is defined as:
\begin{equation}
 TS = -2\log\frac{P(Data|\mathcal{H}_{0})}{P(Data|\mathcal{H}_{1})}.
\end{equation}

A larger \textit{TS} value indicates a smaller compatibility with background. The probability of observing the data given a certain hypothesis, as written in the \textit{TS} definition, is given by a likelihood function. The likelihood function contains parameters that are fitted in the analysis and is defined as: 

\begin{equation}
\label{eq:likelihood}
 \mathcal{L}(\vec{x}_s, n_{s}, \gamma, ...) = \prod_{i=0}^N(\frac{n_{s}}{N}S_{i}+(1-\frac{n_{s}}{N})B_{i})
\end{equation}
where $i$ runs over the number of events $N$, $\vec{x}_s$ is the source position, $n_s$ is the number of signal events, $\gamma$ is the source spectral index and $S_i$ and $B_i$ are respectively the signal and the background probability density functions (PDFs). 

The signal PDF is composed of a spatial, energy and time PDF
\begin{equation}
\label{eq:signalPDF}
 S_{i} = P_{i}^{sig}(\sigma_{i},\vec{x}_{i}|\vec{x}_{s})\cdot\epsilon_{i}^{sig}(E_{i},\delta_{i}|\gamma)\cdot T_{i}^{sig}, 
\end{equation} where $P_{i}^{sig}$ is the spatial PDF and $\epsilon_{i}^{sig}$ is the energy PDF. They are described in \citet{Aartsen_2013}. The time PDF $T_{i}^{sig}$ is specific to each analysis and will be described in the corresponding sections. The background PDF
\begin{equation}
\label{eq:backgroundPDF}
 B_{i} = P_{i}^{bkg}(\delta_{i})\cdot\epsilon_{i}^{bkg}(E_{i},\delta_{i})\cdot T_{i}^{bkg}
\end{equation}
 has the same structure as the signal PDF and is decribed in  \citet{Aartsen_2013}.
 
The background spatial PDF $P_{i}^{bkg}(\delta_{i})$ gives the density of atmospheric neutrinos and muons with respect to the declination $\delta_i$ of the event $i$. It is uniform in right ascension when we consider data over a sufficiently long period. For short time scales, of the order of less than a day, a dependency in right ascension appears. This is caused by the fact that incoming neutrinos aligned with the strings of the detector are better reconstructed. This effect fades away with Earth rotation. 
In order to treat correctly short time scales, the right ascension-independent PDF is multiplied by a local coordinate distribution, $P_{i}^{bkg}(\phi_{i},cos\theta_{i})$, which depends on both the zenith $\theta$ and the azimuth $\phi$ angle for each event $i$ \citep{Christov:2016zti}. It is to be noticed that at the South Pole the relation between the zenith $\theta_i$ and the declination $\delta_i$ is largely simplified to $\delta_i = -90^\circ + \theta_i$. This is important because it means that the background variation at the local detector level (zenith) is equivalent to its variation in equatorial coordinates (declination).

The energy PDF gives the probability of finding an event with measured energy $E_i$ coming from the atmospheric background at the declination of the event $\delta_i$.
$E_i$ is an energy proxy of the true neutrino energy based on the visible muon energy in the detector.

Finally, the time-dependent term of the background PDF is taken to be constant for the background even though some seasonal variations are observed in the atmospheric event rates due to changes of the atmosphere density \citep{abbasi2011time}. These variations are at the level of $\pm 15\%$ of the rate of down-going muons, but are much reduced for up-going neutrinos coming from all latitudes across the Earth.

\subsection{Time PDF of the all-sky scans}

The time-dependent term, $T_i^{sig}$ is defined by a Gaussian function, with width $\sigma_T$ and mean $T_0$:
\begin{equation}
    T_i^{sig} = \frac{1}{\sqrt{2\pi}\sigma_T}e^{\left(-\frac{(t_i-T_0)^2}{2\sigma^2_T}\right)},
\end{equation}
where $t_i$ is the arrival time of the $i^{th}$ event. We tested also step functions to represent a flare, but the results do not change in a significant way.

In order to avoid an undesired bias of the fit towards short flares in a specific time period, a penalizing term $T/\sqrt{2\pi}\sigma_T$ is added to the original test statistics, as explained in \citet{BRAUN2010175}:
\begin{equation}
\label{eq:TS_allskyscan}
    TS = -2\log \left[ \frac{T}{\sqrt{2\pi} \hat{\sigma}_T} \times \frac{\mathcal{L}(n_s = 0) }{\mathcal{L}(\hat{n}_s, \hat{\gamma}_s, \hat{\sigma}_T, \hat{T}_0)} \right]
\end{equation}
where $\hat{n}_s$, $\hat{\gamma}_s$, $\hat{\sigma}_T$, $\hat{T}_0$ are the best-fit values of the parameters defined above. 

The all-sky scan is realized by creating a $0.1^\circ{} \times 0.1^\circ{}$ grid covering the whole sky, for declinations between $-85^\circ{}$ to $+85^\circ{}$ (not enough statistics are collected at the poles to apply the analysis there) and by fitting the free parameters in order to maximize the likelihood at each point. This results in a \textit{TS} value at each point of the sky grid. 

This analysis is performed to determine the hottest spots in the Northern and in the Southern sky. In order to calculate the post trial values of the \textit{TS} for the hottest spots, we need to account for all the directions looked at. Because the sample is background-dominated, we use `pseudo-experiments' (trials) containing only background created by assigning a random location in the sky (right ascension) to each event \citep{Aartsen:2015wto}. By applying this procedure, which preserves the distributions in detector coordinates\footnote{This is because the distribution of events is expected to be flat in right ascension due to the continuous flux of atmospheric neutrinos and muons, the Earth rotation and the fact that the acquisition is continuous and has no stop at a fixed time every day.}, we make sure that no real signal, that could potentially be contained in the sample, remains in the trial sample. 

\begin{figure}[h]
\begin{center}
\begin{minipage}{0.49\linewidth}
\includegraphics[width=.9\textwidth]{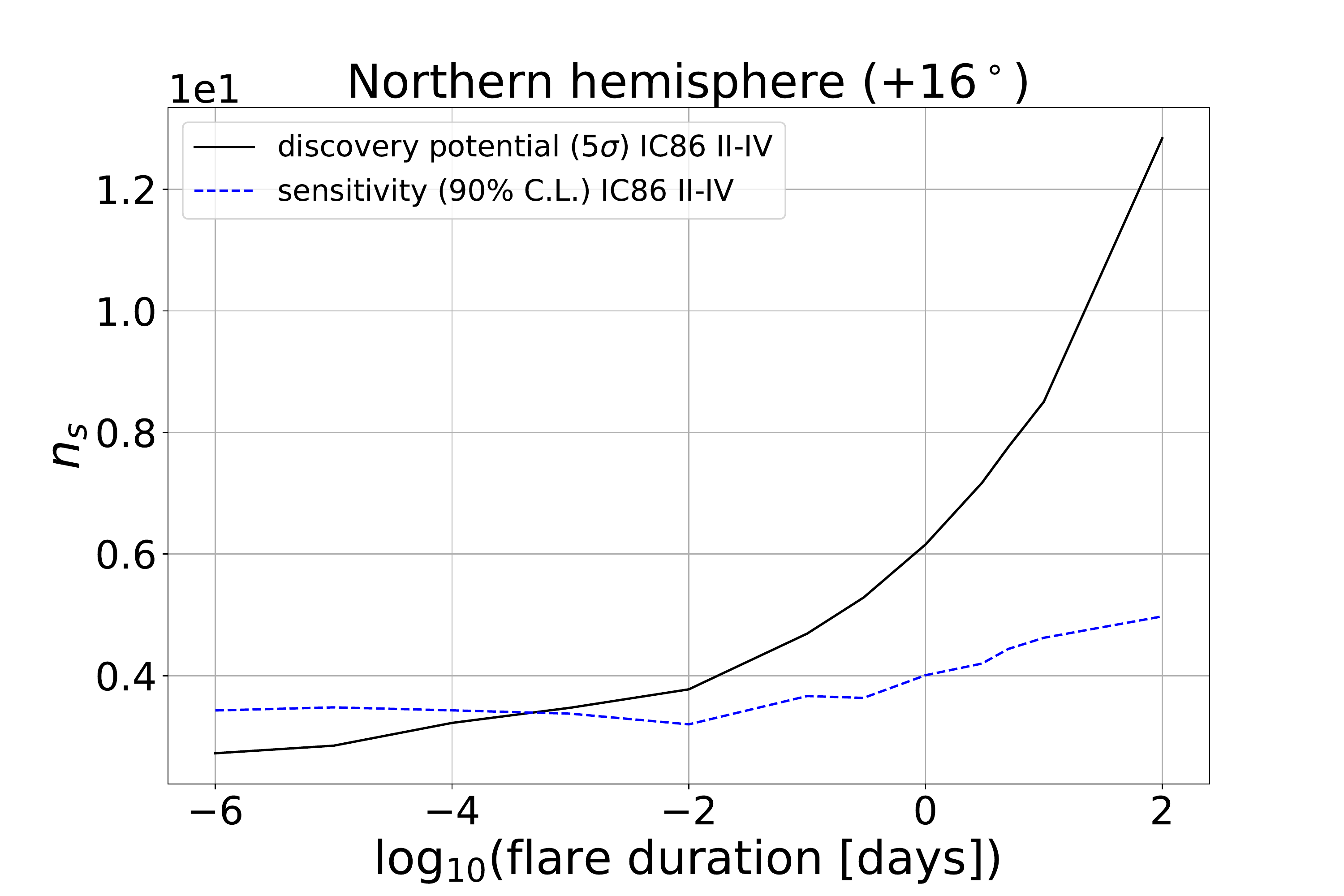}\\ 
\includegraphics[width=.9\textwidth]{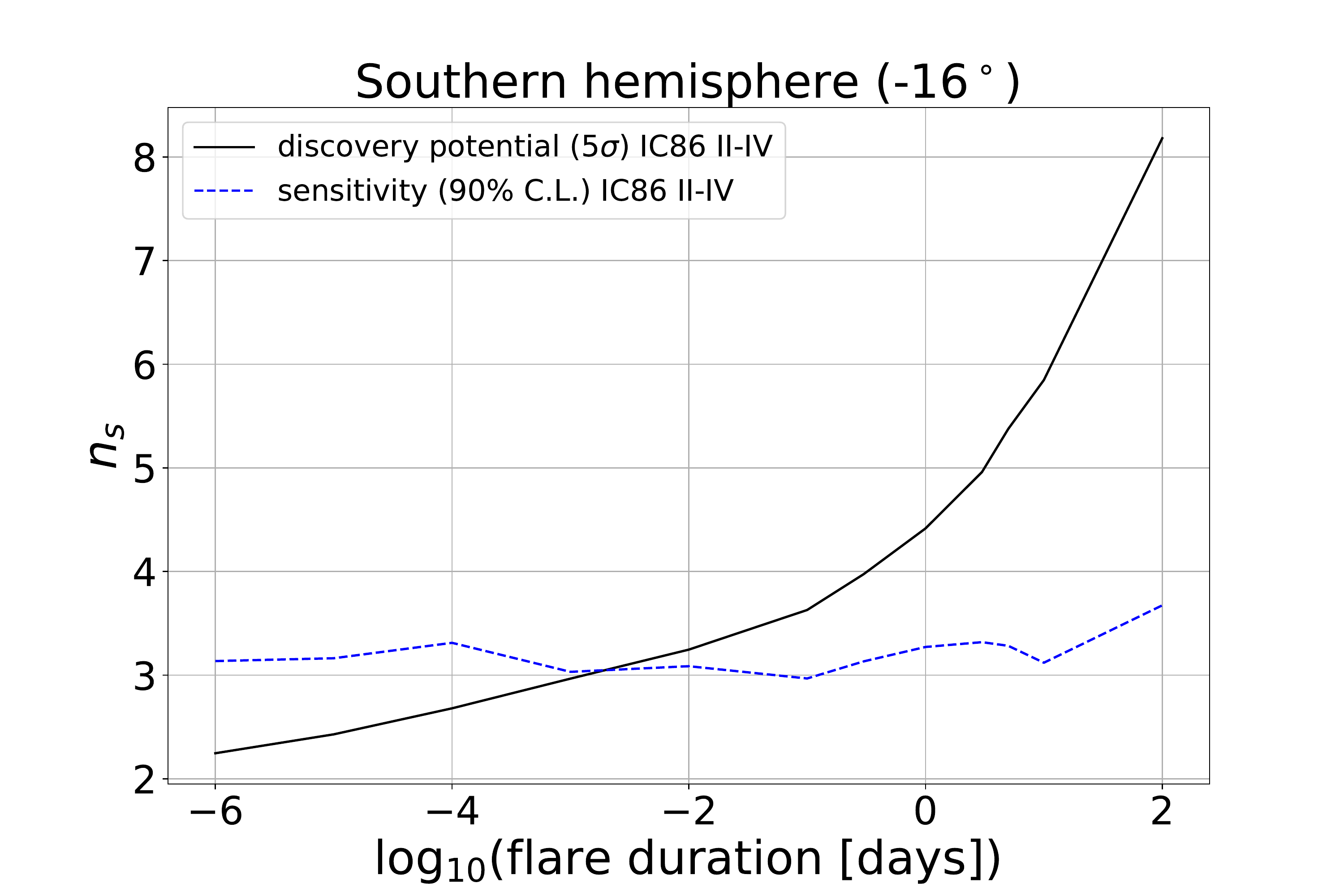}\\ 
\end{minipage}
\hfill
\begin{minipage}{0.49\linewidth}

\includegraphics[width=.9\textwidth]{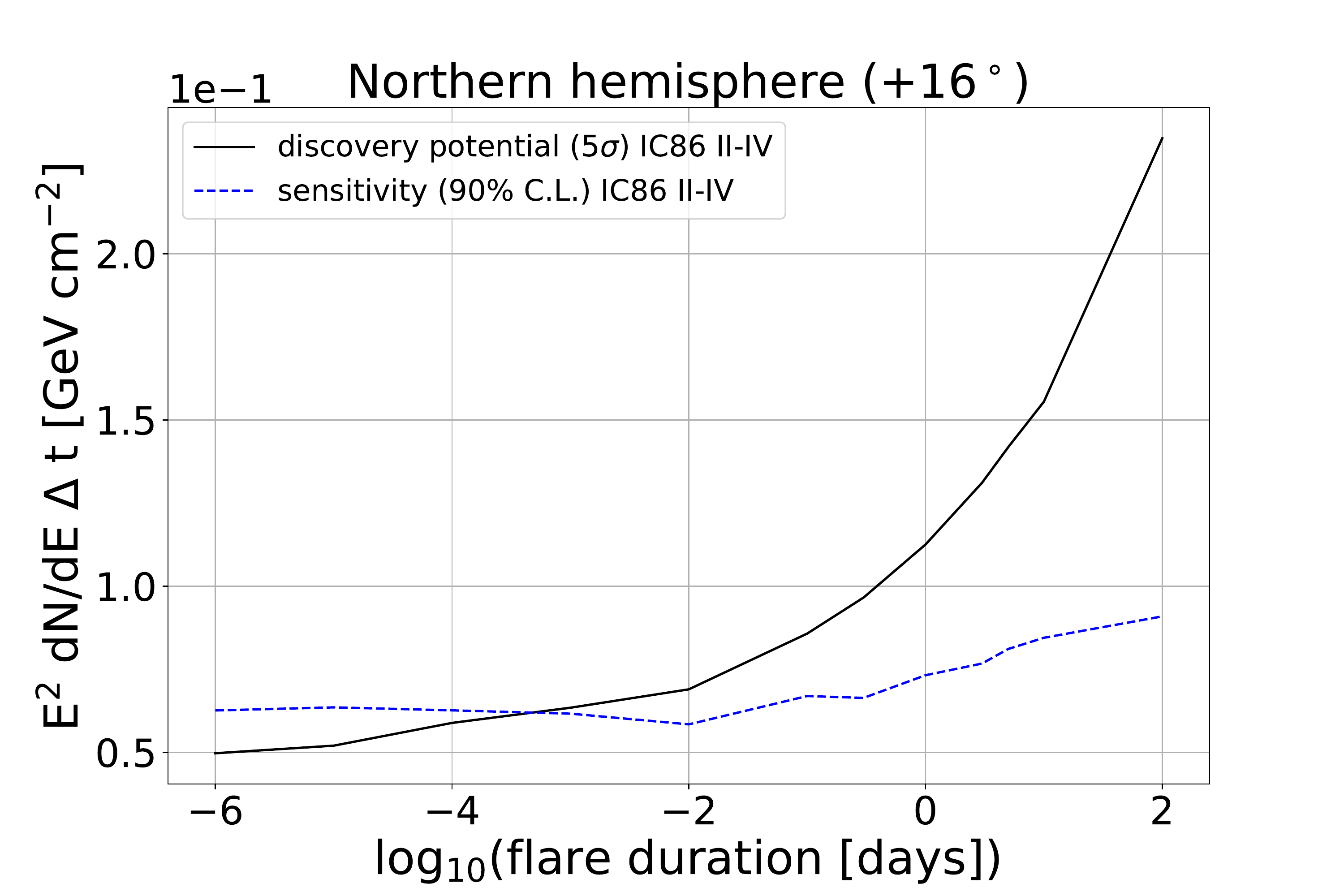}\\ 
\includegraphics[width=.9\textwidth]{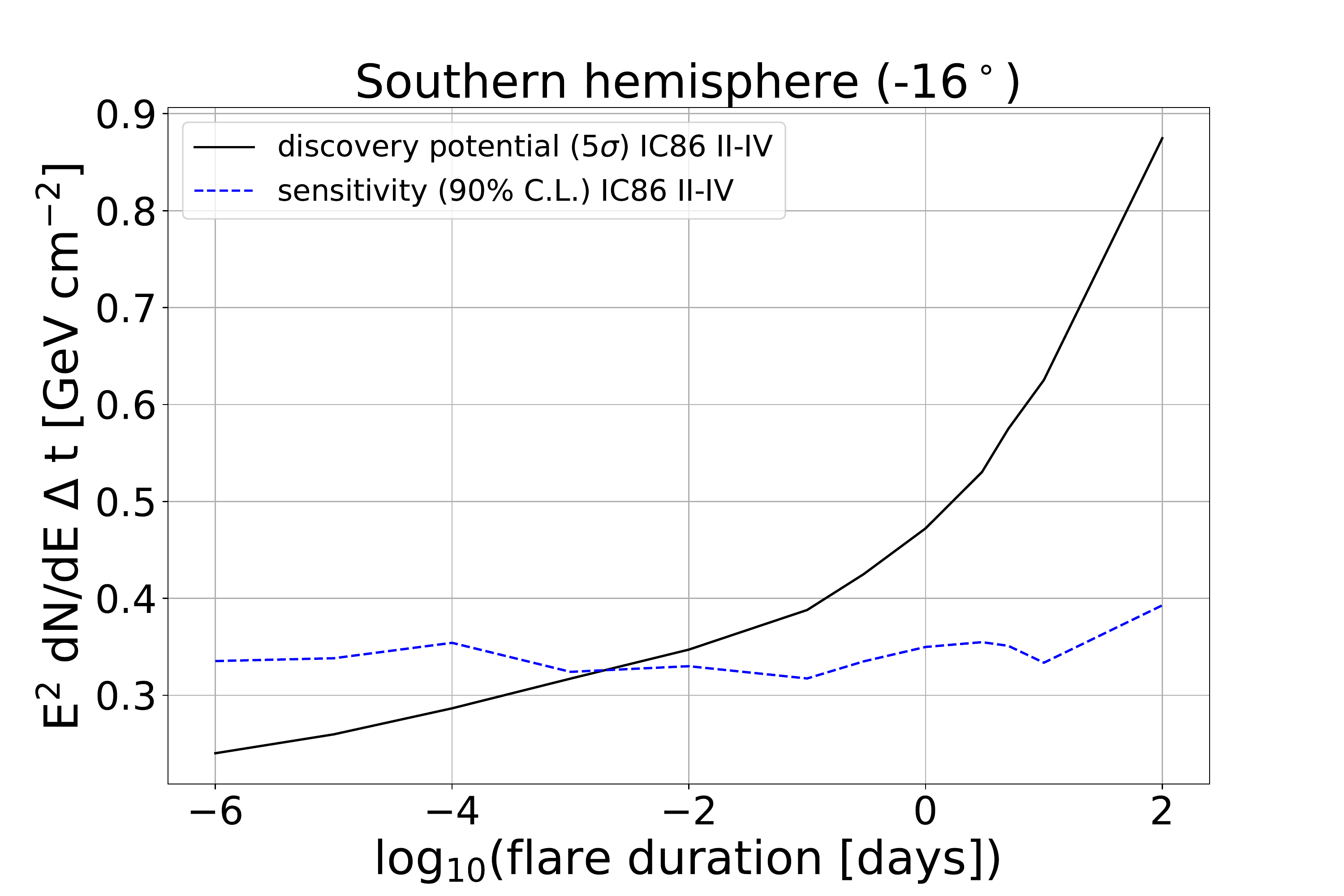}\\ 
\end{minipage}
\caption{The 5$\sigma$ discovery potential (solid black line) and the sensitivity (dashed blue line) for IC86 II-IV shown in terms of the mean number of signal events ($n_s$) (left) and in terms of the time-integrated flux (right) for a fixed source in the Northern hemisphere at +16$^\circ$ in declination (top) and in the Southern hemisphere at -16$^\circ$ (bottom) with an $\rm E^{-2}$ spectrum.
}
\label{fig:dp_sensi_2012_2015}
\end{center}
\end{figure}

\begin{figure}[h]
\begin{center}
\begin{minipage}{0.49\linewidth}
\includegraphics[width=.9\textwidth]{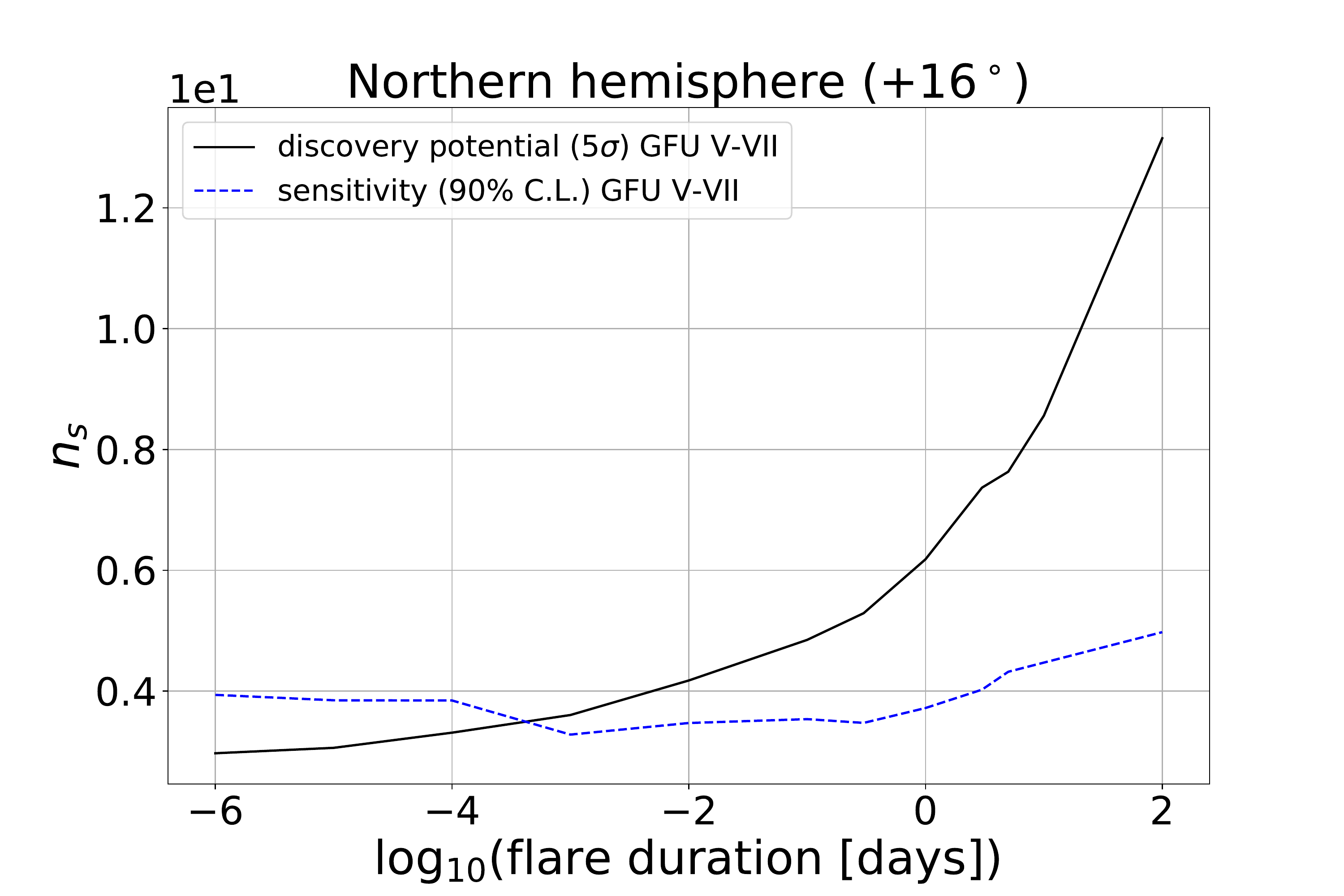}\\ 
\includegraphics[width=.9\textwidth]{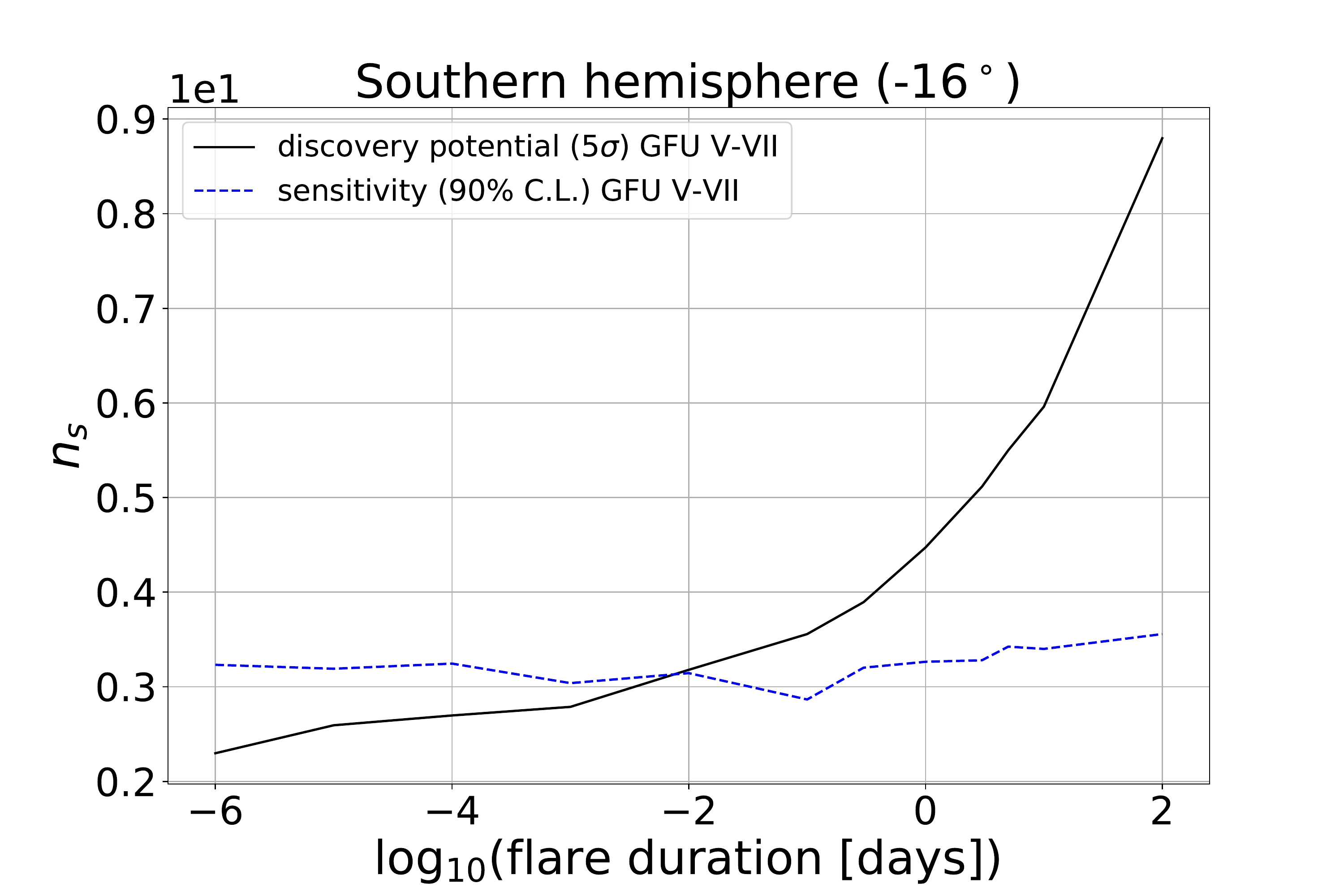}\\ 
\end{minipage}
\hfill
\begin{minipage}{0.49\linewidth}

\includegraphics[width=.9\textwidth]{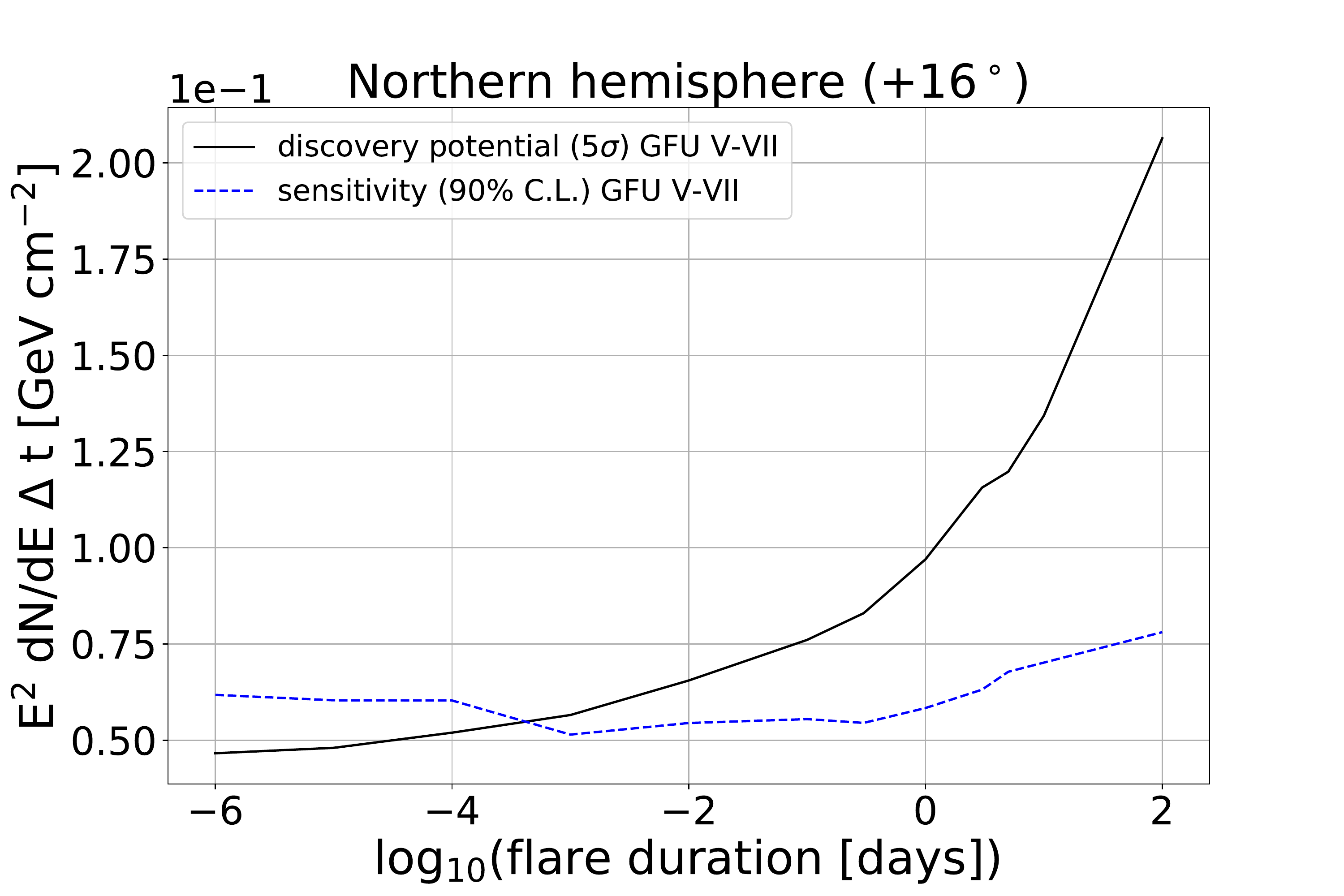}\\ 
\includegraphics[width=.9\textwidth]{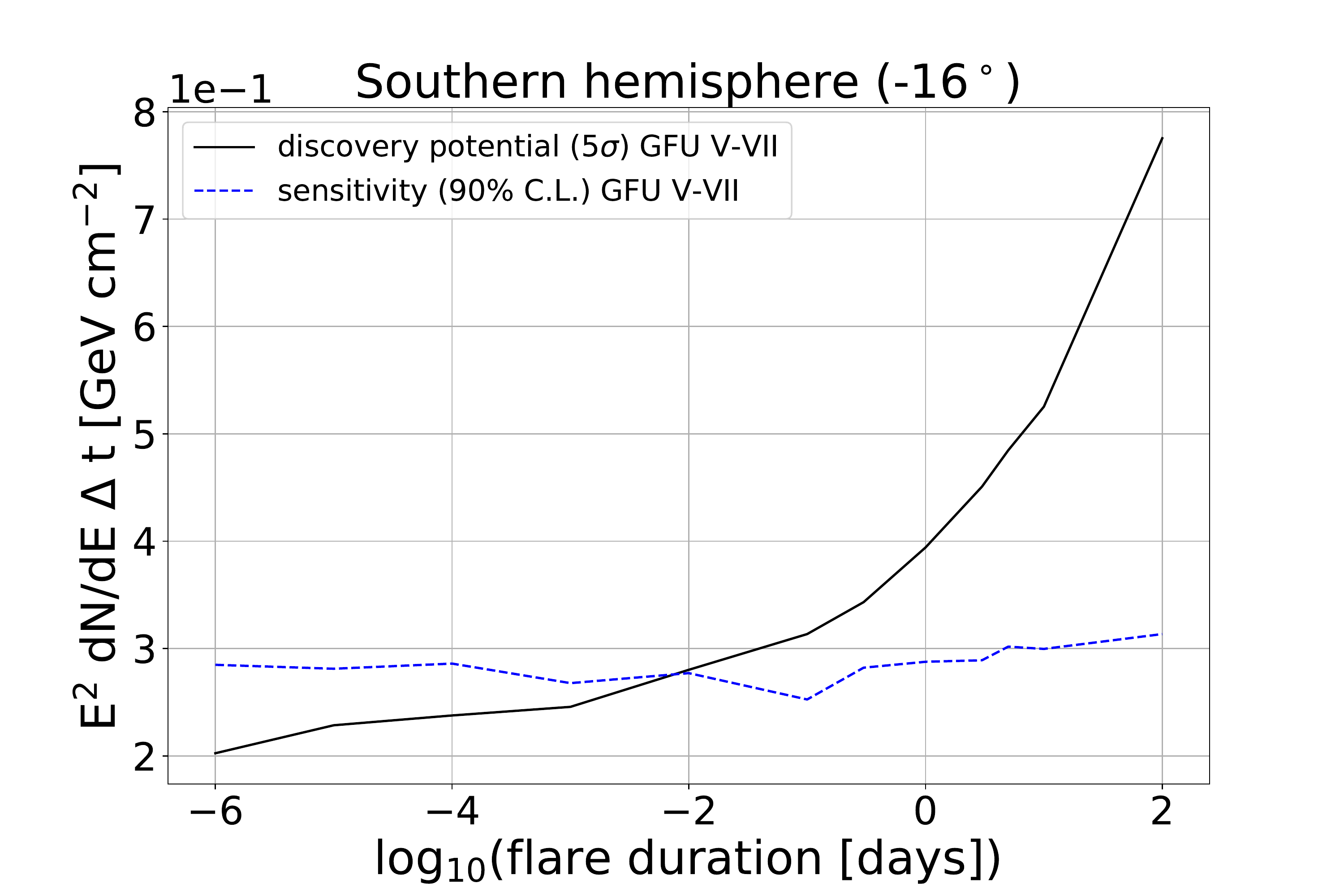}\\ 
\end{minipage}
\caption{The 5$\sigma$ discovery potential (solid black line) and the sensitivity (dashed blue line) for IC86 V-VII shown in terms of the mean number of signal events ($n_s$) (left) and in terms of the time-integrated flux (right) for a fixed source in the Northern hemisphere at +16$^\circ$ in declination (top) and in the Southern hemisphere at -16$^\circ$ (bottom) with an $\rm E^{-2}$ spectrum.
}
\label{fig:dp_sensi_2015_2017}
\end{center}
\end{figure}

The performance of the analysis can be estimated in terms of the sensitivity, defined as the average number of signal events ($n_s$) required so that 90\% of the generated `equivalent' experiments have a \textit{TS} greater than 50\% of the background trials (i.e. where no signal events are injected on top of the background events), and in terms of the discovery potential, defined as the $n_s$ required so that 50\% of the trials have a \textit{TS} greater than $5\sigma$ of the background distribution, i.e. a $p$-value of $3\times10^{-7}$.
The average number of events observed from a source is directly related to its flux. The usual way of representing it in time-dependent analyses is through the time-integrated flux or through the fluence, which is the energy-flux integrated over the time of the selected flare and over the energy interval, limited to the 5\% and the 95\% energy percentiles of the event sample \citep{Aartsen:2015wto}.
Sensitivity and discovery potential in terms of $n_s$ and time-integrated flux are shown in Figs.~\ref{fig:dp_sensi_2012_2015} and \ref{fig:dp_sensi_2015_2017} as a function of the logarithm of the duration of the flare for the periods from 2012 to 2015 (IC86 II-IV) and from 2015 to 2017 (IC86 V-VII), respectively. 


\subsection{Time PDF of 3C 279 flare analysis}
\label{sec:3C279FlareAnaMethod}

The 3C 279 flare analysis is a \textit{triggered} time-dependent point-source analysis, i.e. its search time window is determined by the flaring behavior of the source in gamma rays. To build such an analysis, the gamma-ray information is combined together with the neutrino information.  

\begin{figure}[h!]
\begin{center}
\includegraphics[scale=.6]{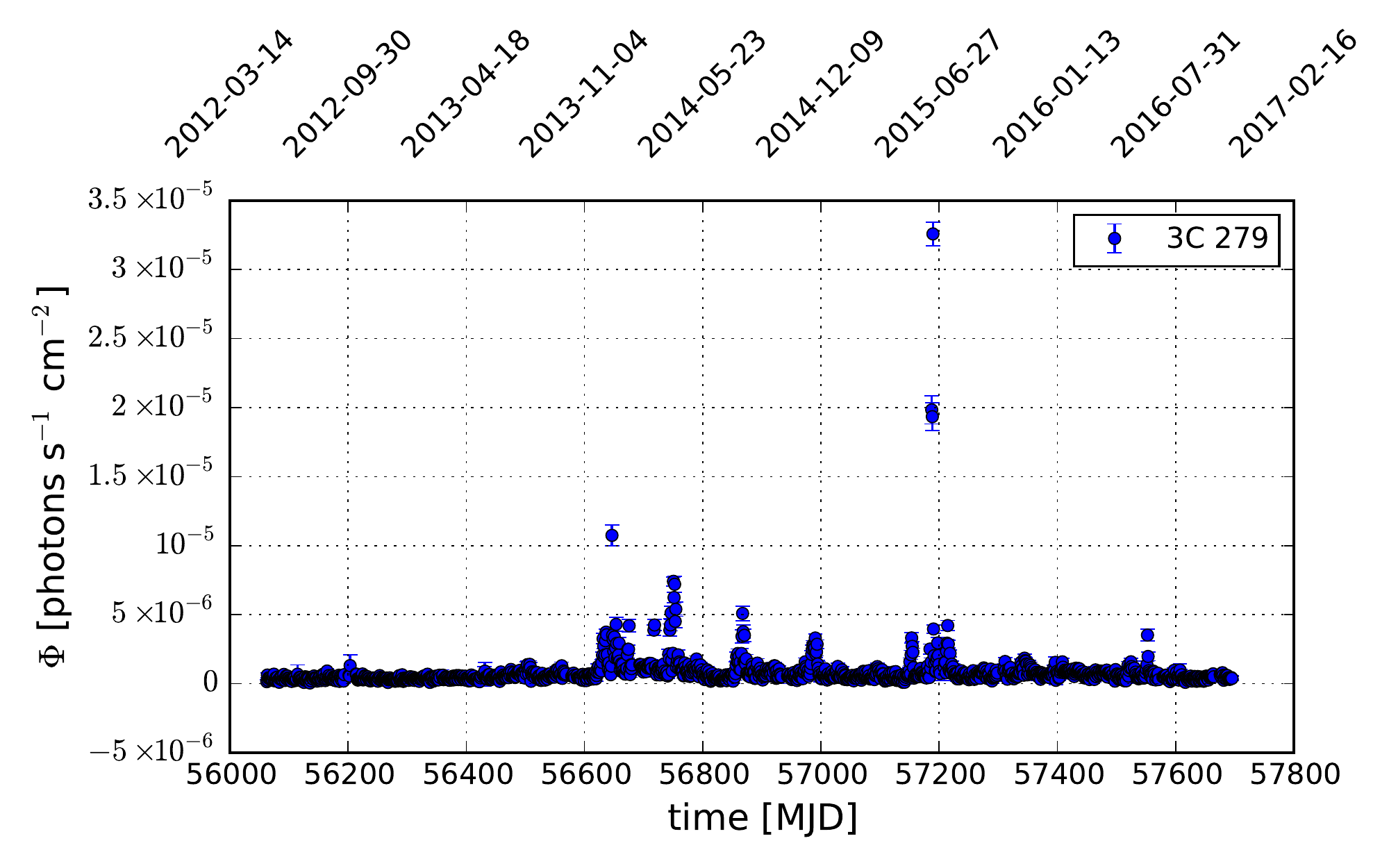}
\end{center}
\caption{Integral photon flux above 100~MeV of the 3C 279 blazar with one-day binning, produced using aperture photometry. The data come from \textit{Fermi}-LAT.}
\label{fig:lc}
\end{figure}

The gamma-ray emission of 3C 279 is recorded by \textit{Fermi}-LAT and its light curve, shown in Fig.~\ref{fig:lc}, is produced through an aperture photometry analysis\footnote{Because of its fast execution and simplicity, this analysis method was preferred to a full-likelihood approach, despite the better flux precision of the latter, as the resulting light curve is only used in the time-dependent analysis for the variability of its relative flux (see Sec.~\ref{sec:3C279FlareAnaMethod}).} using a one-day binning. With the help of the analysis tools provided by the \textit{Fermi}-LAT collaboration (\textit{Fermi Science Tools v10r0p5} package\footnote{\url{https://fermi.gsfc.nasa.gov/ssc/data/analysis/documentation/}}), open access photometric data from \textit{Fermi}-LAT are downloaded. We use photons of the P8R3$\_$SOURCE class with front and back conversions \citep{Atwood:2013rka}. Those photons are then selected within $2^\circ{}$ from the source, in order to further reduce the background. Photon events with zenith angles greater than $90^\circ{}$ are excluded to avoid contamination due to the Earth's albedo. The photon energy range goes from 100~MeV to 500~GeV as recommended by the \textit{Fermi}-LAT collaboration.

\begin{figure}[H]
\begin{center}
\includegraphics[scale=.8]{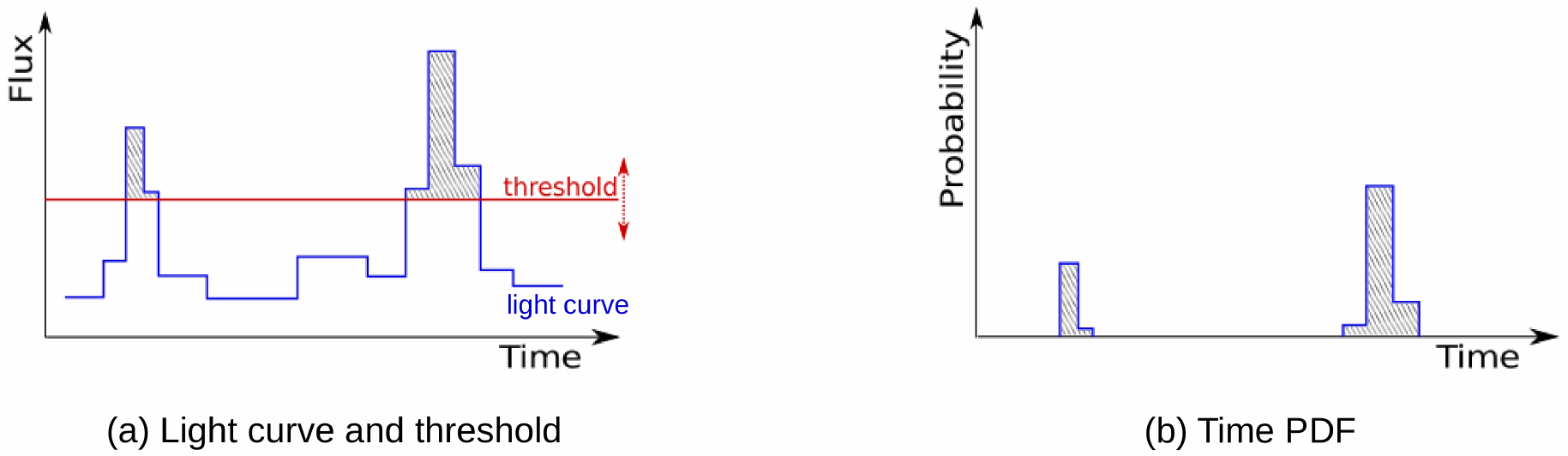}
\end{center}
\caption{(a) The threshold (red) is varied within the flux range defined by the denoised gamma-ray light curve (blue). It defines a region where the source denoised gamma-ray lightcurve (blue) indicates a high-state. (b) The time PDF is constructed from the light curve and the value of the threshold.}
\label{fig:tpdf}
\end{figure}

After the light curve is created, it is de-noised using the Bayesian block method \citep{scargle2013studies,resconi2009classification}. It allows to remove the non-representative fluctuations of the emission and to keep only the main features of the light curve. The light curve is modeled into blocks, each meant to represent a Poisson distribution with a constant mean, extending for a certain duration. The parameter of the Bayesian blocks, called $F_B$, was set to 5, following a previous study on similar IceCube data \citep{Aartsen:2015wto}. A representation of a light curve turned into Bayesian blocks can be seen in Fig. \ref{fig:tpdf} (a).

The gamma-ray and neutrinos data are analyzed over a period of 11 days, from 2015 June 11 to 2015 June 22, which is chosen to isolate the bright flare and whose exact duration was selected based on a series of tests, aiming at maximizing the 5$\sigma$ discovery potential (50\% C.L.). 

The likelihood function of this analysis follows the general form defined in Eq.~\ref{eq:likelihood}. The background PDF is described by Eq.~\ref{eq:backgroundPDF}. The signal spatial and energy PDFs are defined in Eq.~\ref{eq:signalPDF}, while the signal time PDF is specific to the triggered analysis. It depends on two parameters, fit in the likelihood maximization: the time lag, $D_t$, and the threshold, $f_{th}$. 

The time lag describes a shift of the entire light curve in time in the range of plus or minus half a day. This tolerance accounts for binning effects, when, for example, a photon flare is spread between contiguous one-day bins and not fully contained within them. While this is relevant for a light curve containing a single flare, such a binning effect would be averaged out when dealing with multiple flares.

If we assume that the gamma rays recorded in the light curve are produced via hadronic interactions of cosmic rays, a higher astrophysical neutrino flux is expected where a higher gamma-ray flux is detected. However, the neutrino background should stay constant. This provides an opportunity to restrict the search for astrophysical neutrinos to the periods of enhanced gamma-ray activity, thus maximizing the signal while reducing the background. The neutrino signal enhancement might be even larger during flares, since in many models of neutrino emission, the neutrino-production efficiency is actually doubly enhanced during flares, due to the fact that both the proton injection and the target photon field are expected to be enhanced at the same time \citep{Oikonomou_2019}.

The flux level at which we consider the object to be in a flaring state is not well defined. The threshold, which is a parameter of the fit, determines where the source is considered in flaring state since it results from the likelihood method optimisation of the signal over background. The construction of the signal time PDF from the variation of the threshold is illustrated by Fig.~\ref{fig:tpdf}. Below the threshold, the time PDF is defined as zero. Above the threshold, the time PDF is defined by the area of the light curve normalized to one, as shown in Fig.~\ref{fig:tpdf} (b). The value of the threshold is varied during the likelihood maximization. 

The test statistic is defined as 
\begin{equation}
    TS = -2\log \left[ \frac{\mathcal{L}(n_s = 0) }{\mathcal{L}(\hat{n}_s, \hat{\gamma}_s, \hat{D}_t, \hat{f}_{th})} \right]
\end{equation}
with $\hat{n}_s$, $\hat{\gamma}_s$, $\hat{D}_t$, $\hat{f}_{th}$ the best-fit values for the number of signal events, the spectral index, the time lag and the threshold of the time PDF, respectively. Notice that in the fit the parameter $\gamma_s$ is allowed to vary between 1 and 4.

\begin{figure}[hbt]
\begin{center}
\centering \includegraphics[scale=0.5]{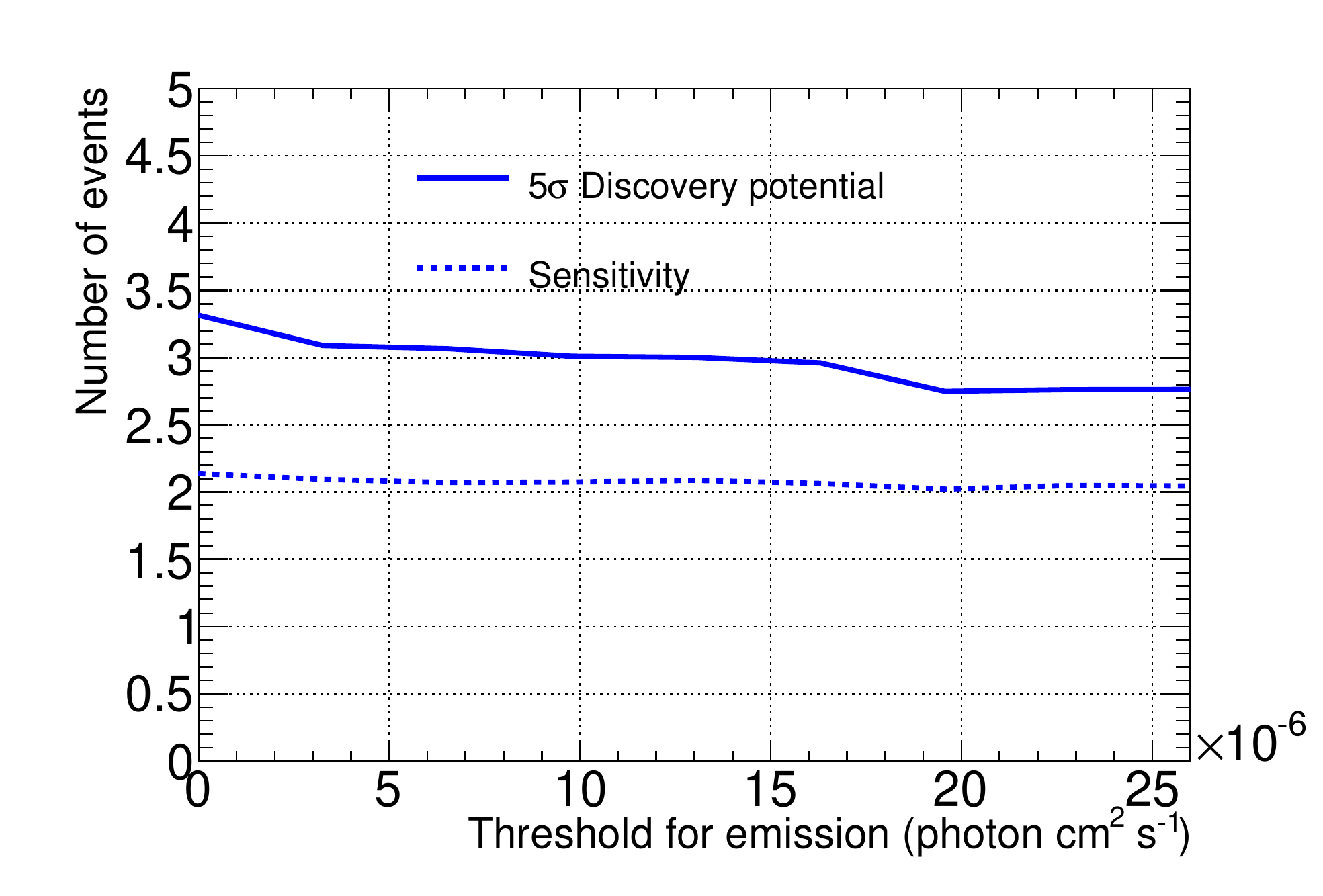}
\caption{Sensitivity and 5$\sigma$ discovery potential for the 11-day flare analysis in number of neutrino signal events as a function of the gamma-ray threshold level in daily bins.}
\label{fig:3C_279_dp_sensi_limits}
\end{center}
\end{figure}

The $5\sigma$ discovery potential and the sensitivity of this analysis, shown in Fig.~\ref{fig:3C_279_dp_sensi_limits}, are functions of the value of the threshold, $f_{th}$, given in terms of flux. The discovery potential and the sensitivity are largely insensitive to the threshold, even though there is a small tendency for them to improve with larger threshold. The reason for this is that the higher the threshold is, the more the time PDF is restricted to a short period and the less background is included in the fit. Logically, fewer events are needed in order to make a discovery in such case. An average of 3 events is required for a discovery while about 2 events are needed  to reach the 90\% sensitivity level. 

\section{Results}
\subsection{Time-dependent all-sky scans}
\subsubsection{IC86 II-VI analysis}

\begin{figure}[!ht]
\begin{center}
\includegraphics[width=.7\textwidth]{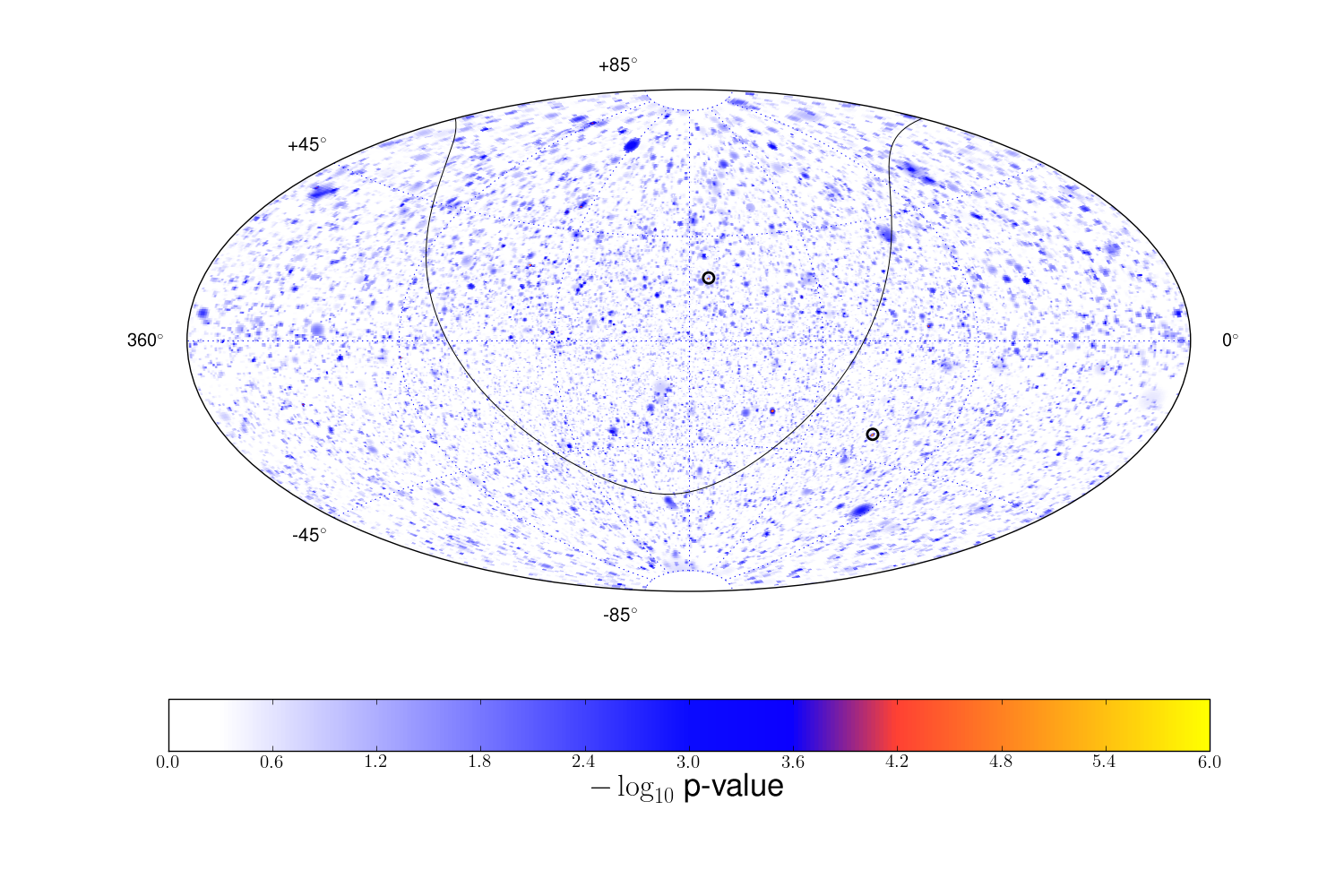}
\end{center}
\caption{IC86 II-IV skymap in equatorial coordinates showing the pre-trial $p$-values for the best-fit flare hypothesis tested in each direction of the sky. The strongest time-dependent Gaussian-like signal was found in the Northern sky at (RA, Dec) = (170.4$^{\circ{}}$, 28.0$^{\circ{}}$), with post-trial $p$-value of 17.7\%. The solid black curve indicates the Galactic plane and the hottest spot in each hemisphere is circled.}
\label{fig:unblinded_skymap_2012_2015}
\end{figure}

The all-sky scan is realized by running the fit, resulting in a \textit{TS} (Eq.~\ref{eq:TS_allskyscan}), at each point of a $0.1^\circ{}$ x $0.1^\circ{}$ grid over the sky (except the poles). Each \textit{TS} is converted into a $p$-value, assuming that the \textit{TS} distribution follows a $\chi^{2}$ distribution of three degrees of freedom, even though four parameters are fitted in the analysis. Indeed, due to correlations between parameters and some having limited ranges, a $\chi^2$ distribution of three degrees of freedom was found to be the one best-representing the \textit{TS} distribution. Fig.~\ref{fig:unblinded_skymap_2012_2015} shows a sky map of the entire sky in equatorial coordinates for the period from 2015 to 2017 and reports the $p$-value found at each point of the grid.

\begin{figure}[!ht]
\begin{center}
\begin{minipage}{0.39\linewidth}

\includegraphics[width=1\textwidth]{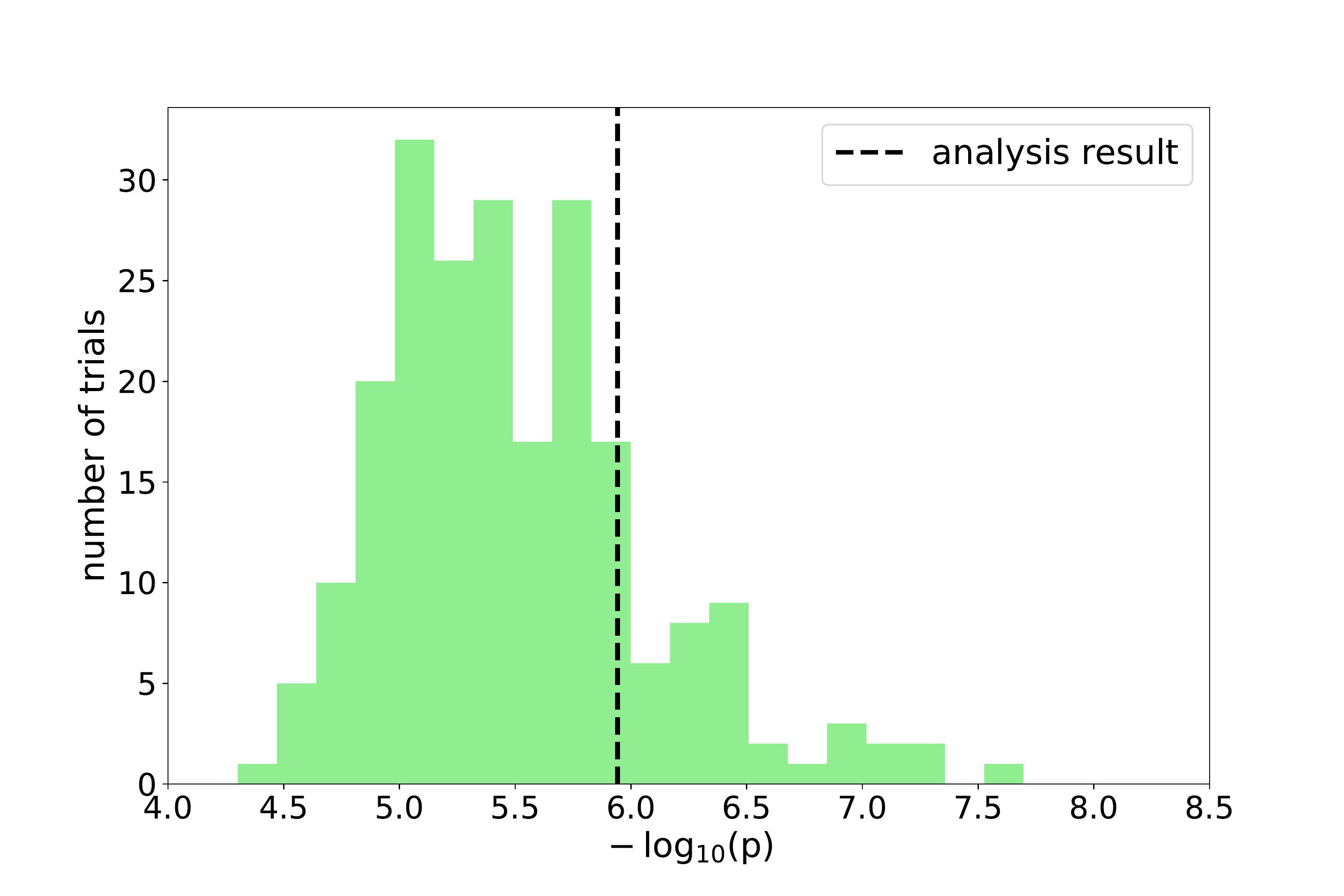}\\ 
\end{minipage}
\hfill
\begin{minipage}{0.59\linewidth}

\includegraphics[width=1.1\textwidth]{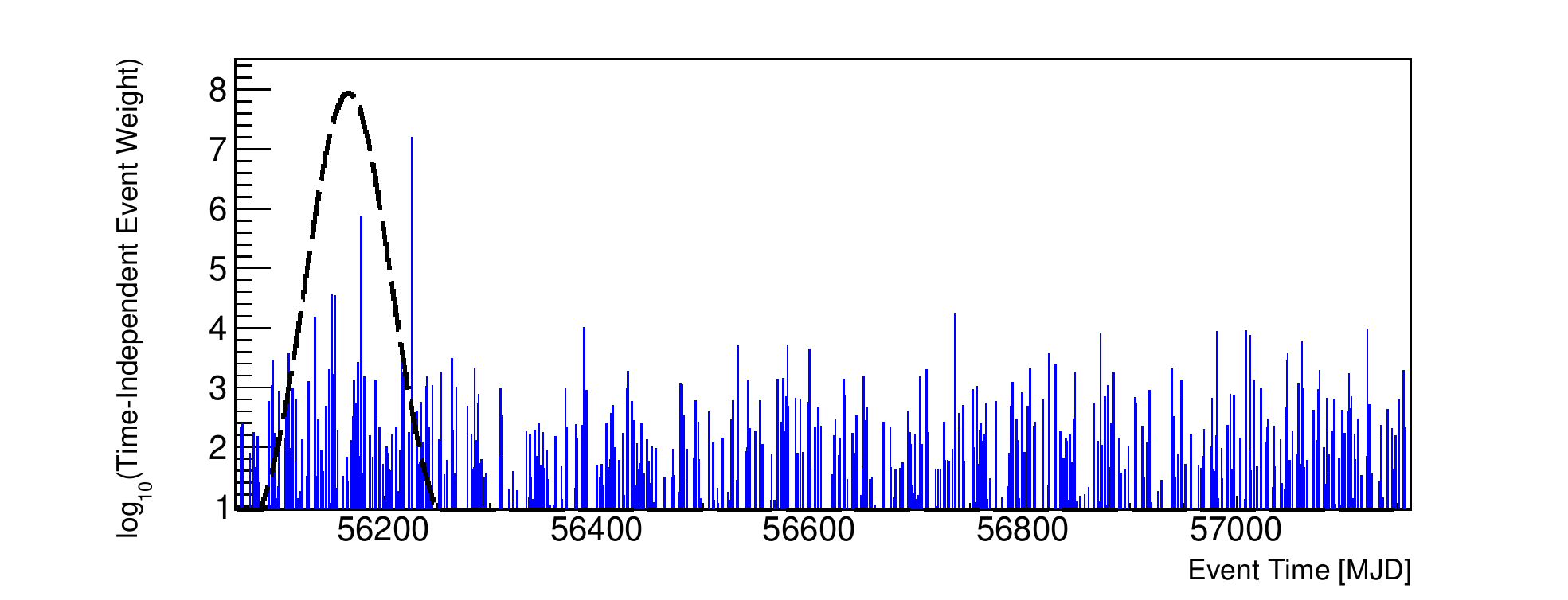}\\ 
\end{minipage}
\caption{Northern hemisphere: Left: expected background $p$-value distribution obtained from scrambled data in green compared to the measured most significant pre-trial $p$-value (shown as the black vertical dashed line) in the Northern sky. The inferred post-trial $p$-value is 17.7\%. Right: the time-independent event weights, evaluated for the IC86 II-IV data in the Northern hemisphere, at a source direction $\vec{x}_s$ defined by the hottest spot (RA, Dec) = (170.4$^{\circ{}}$, 28.0$^{\circ{}}$). The best-fit Gaussian time PDF is shown in black (dashed), with mean $\hat{T}_0$ and sigma $\hat{\sigma}_0$.}
\label{fig:unblinding_result_north_2012_2015}
\end{center}
\end{figure}

\begin{figure}[!ht]
\begin{center}
\begin{minipage}{0.39\linewidth}
\includegraphics[width=1\textwidth]{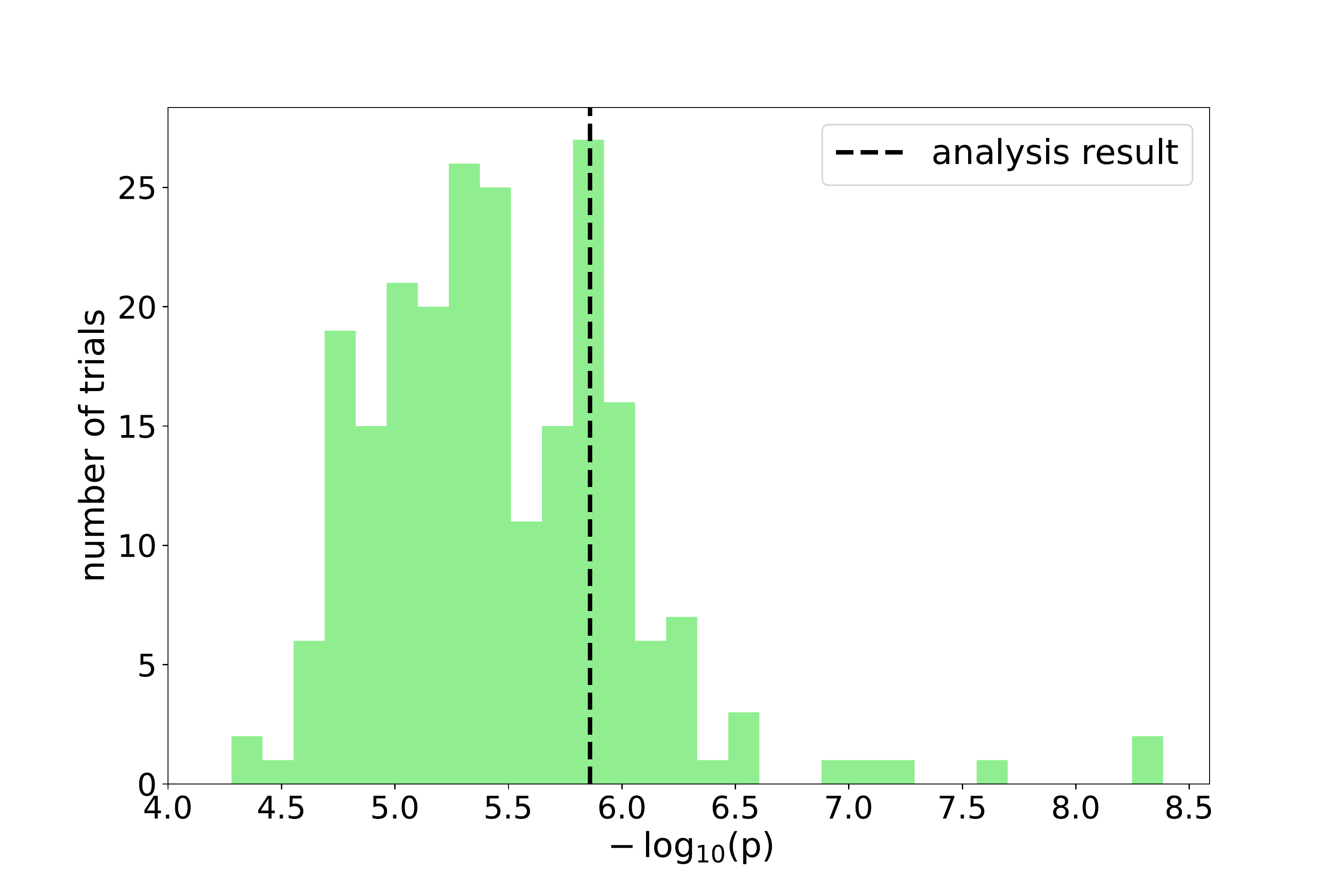}\\ 
\end{minipage}
\hfill
\begin{minipage}{0.59\linewidth}

\includegraphics[width=1.1\textwidth]{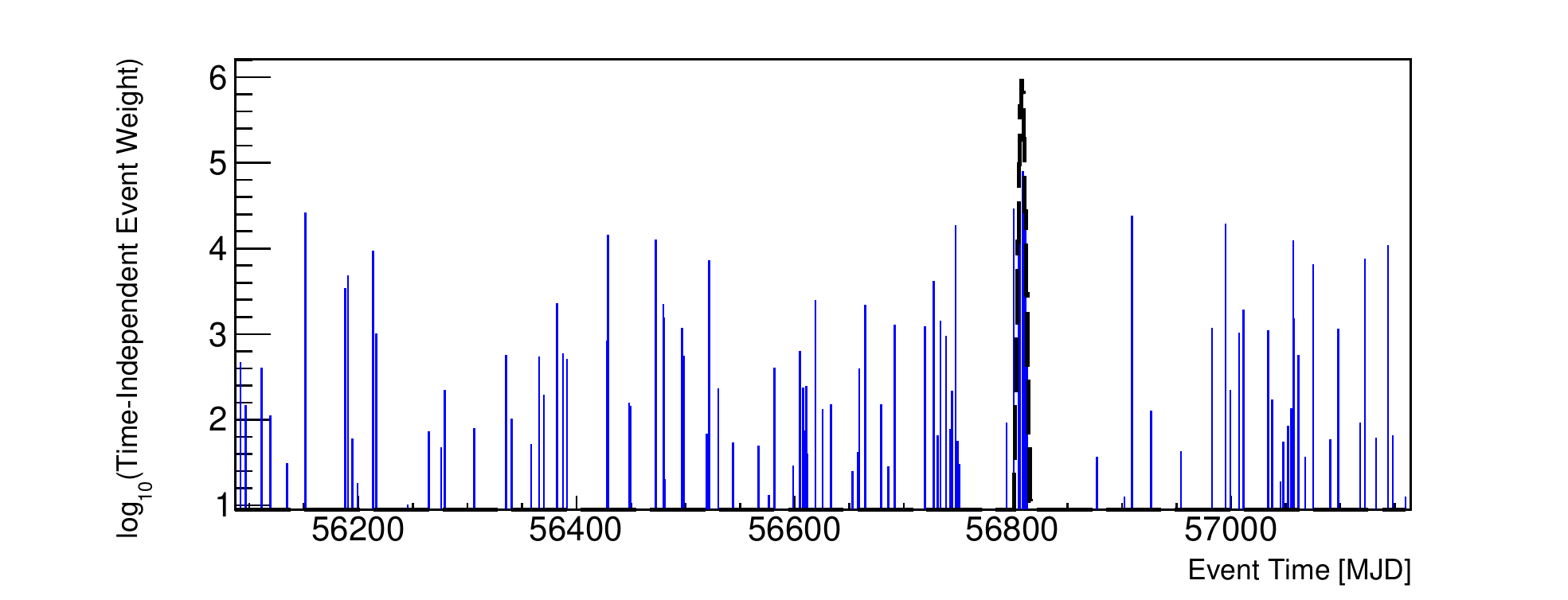}\\ 
\end{minipage}
\caption{Southern hemisphere: Left: expected background $p$-value distribution obtained from scrambled data in green compared to the measured most significant pre-trial $p$-value (shown as the black vertical dashed line) in the Southern sky. The inferred post-trial $p$-value is 24.2\%. Right: the time-independent event weights, evaluated for the IC86 II-IV data in the Southern hemisphere, at a source direction $\vec{x}_s$ defined by the hottest spot (RA, Dec) = (89.45$^{\circ{}}$, -35.95$^{\circ{}}$). The best-fit Gaussian time PDF is shown in black (dashed), with mean $\hat{T}_0$ and sigma $\hat{\sigma}_0$.}
\label{fig:unblinding_result_south_2012_2015}
\end{center}
\end{figure}

The $p$-values shown in the sky map are \textit{pre-trial} $p$-values. The most significant spot in each hemisphere is circled in black on the sky map in Fig.~\ref{fig:unblinded_skymap_2012_2015}. The hottest spot overall was found in the Northern hemisphere at (RA, Dec)=($170.4^\circ{}, 28.0^\circ{}$)\footnote{All equatorial coordinates are quoted for the J2000 epoch.}. The best fit parameters of the Gaussian time PDF are $\hat{T}_0 = 56167.27$~MJD and $\hat{\sigma}_T=40.01$~days. The fitted number of signal events is $\hat{n}_s = 11.79$ and the fitted spectral index is $\hat{\gamma} = 2.11$. The pre-trial significance is $1.38 \times 10^{-6}$, which results in a post-trial $p$-value of 17.7\%. The inferred post-trial $p$-value for the Southern hemisphere is 24.2\%.
These values are compatible with a statistical fluctuation of the background. The observed $p$-value from data together with the distribution of \textit{TS} from pseudo-experiments are shown in the left panels of Fig.~\ref{fig:unblinding_result_north_2012_2015} and \ref{fig:unblinding_result_south_2012_2015}.

In order to help visualize how the different terms of the likelihood contribute to each event, a weight $w_i$ is defined for each event as the ratio between the signal and background PDF of the event without the time term:
\begin{equation}
\label{eq:weights}
w_i = \frac{P_{i}^{sig}(\sigma_{i},\vec{x}_{i}|\vec{x}_{s}) \cdot \epsilon_{i}^{sig}(E_{i},\delta_{i}|\gamma)}{P_{i}^{bkg}(\phi_{i},\theta_{i})\cdot\epsilon_{i}^{bkg}(E_{i},\delta_{i})}
\end{equation}
When the weights are then plotted on the time axis, they immediately allow to visualize which part of the likelihood (spatial/energy or time) dominates the significance.
The right panels of Fig.~\ref{fig:unblinding_result_north_2012_2015} and \ref{fig:unblinding_result_south_2012_2015} show the time-independent weights (Eq.~\ref{eq:weights}) at a source direction $\vec{x}_s$ defined by the hottest spot in the Northern and Southern hemisphere respectively. The best-fit Gaussian time structure with mean $\hat{T}_0$ and sigma $\hat{\sigma}_0$ is overlaid. 

\subsubsection{IC86 V-VII analysis}

\begin{figure}[!ht]
\begin{center}
\includegraphics[width=.7\textwidth]{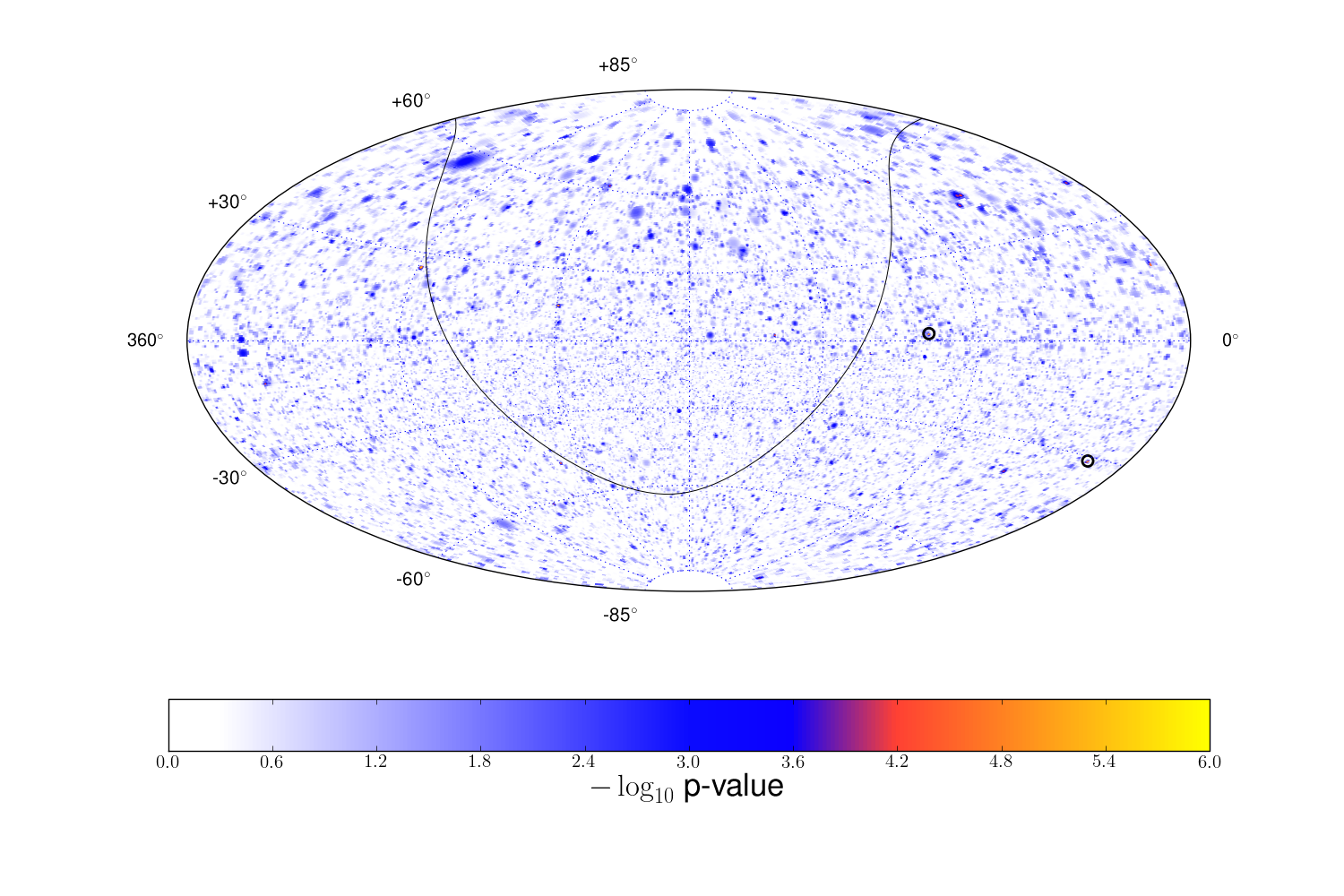}
\end{center}
\caption{IC86 V-VII skymap in equatorial coordinates showing the pre-trial $p$-value for the best-fit flare hypothesis tested in each direction of the sky. The strongest time-dependent Gaussian-like signal was found in the Northern sky at (RA, Dec) = (77.7$^{\circ{}}$ , 2.6$^{\circ{}}$), with post-trial significance of 18.8\%. The solid black curve indicates the Galactic plane and the hottest spots are circled in each hemisphere.}
\label{fig:unblinded_skymap_2015_2017}
\end{figure}

The time period of 2015 to 2017 was treated separately from the previous time period since a different event selection was applied. Fig.~\ref{fig:unblinded_skymap_2015_2017} shows the sky map for the period 2012 to 2015, displaying the pre-trial $p$-values. The most significant value for each hemisphere is circled. The hottest region is found in the Northern hemisphere, like in the previous analysis period, at coordinates (RA, Dec) = ($77.7^\circ{}, 2.6^\circ{}$). The pre-trial $p$-value is $1.3 \times 10^{-6}$, with the best-fit number of signal events being $\hat{n}_s = 25.27$ for a flux with spectral index $\hat{\gamma} = 2.55$ and the most significant time clustering centered at $\hat{T}_0 = 57573.85$~MJD, with a Gaussian width $\hat{\sigma}_0 = 189.6$~days. This results in a post-trial $p$-value of 18.8\%.

\begin{figure}[!ht]
\begin{center}
\begin{minipage}{0.39\linewidth}

\includegraphics[width=1\textwidth]{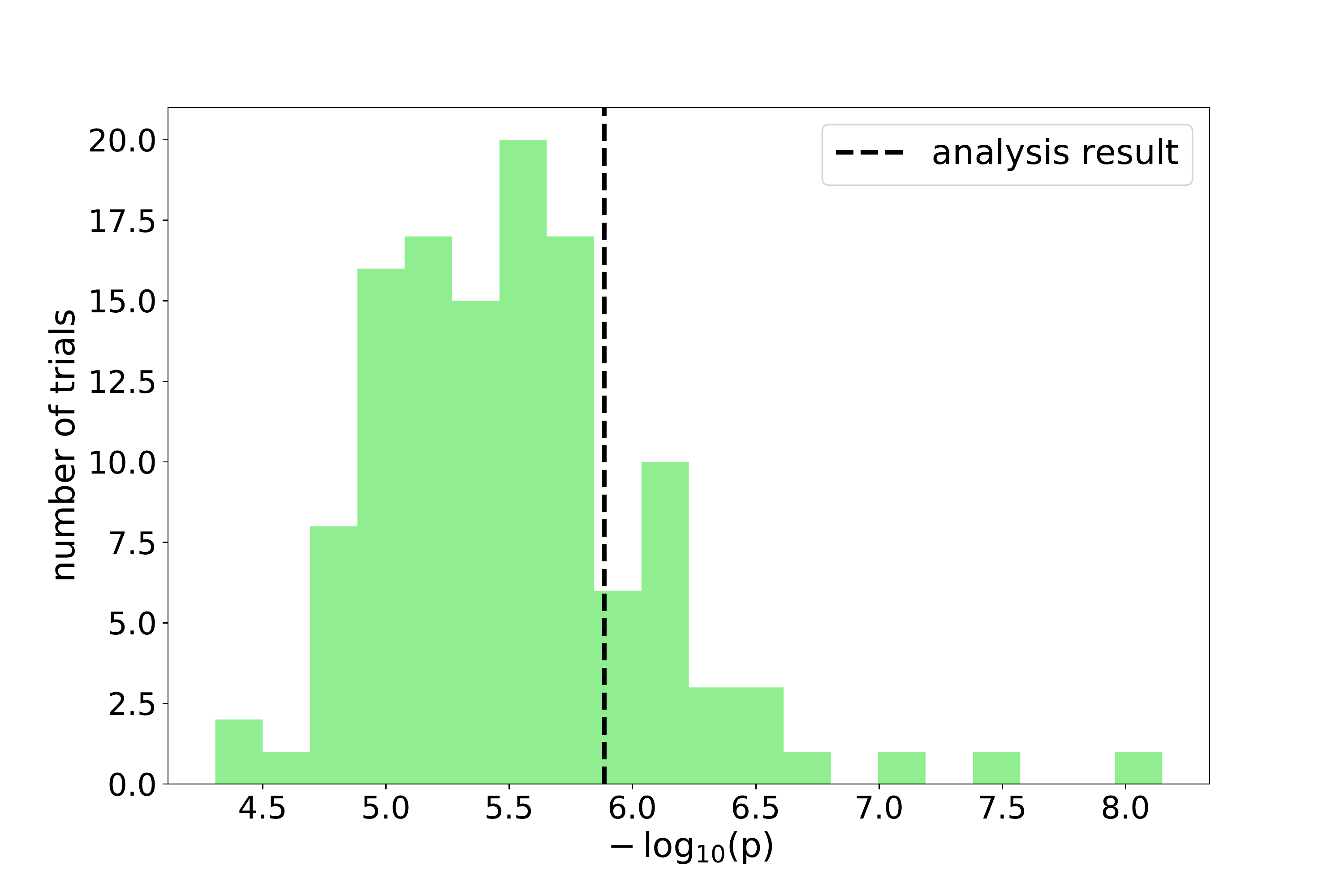}\\ 
\end{minipage}
\hfill
\begin{minipage}{0.59\linewidth}

\includegraphics[width=1.1\textwidth]{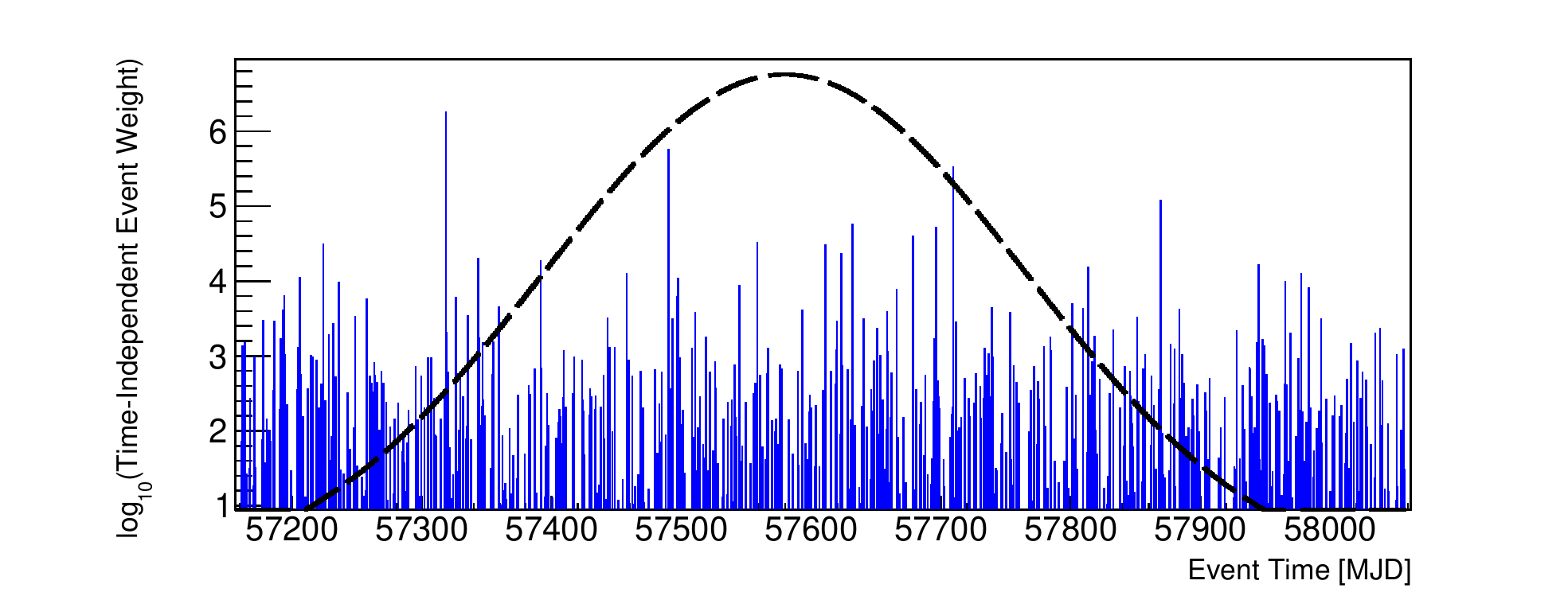}\\ 
\end{minipage}
\caption{Northern hemisphere: Left: expected background $p$-value distribution obtained from scrambled data in green compared to the measured most significant pre-trial $p$-value (shown as the black vertical dashed line) in the Northern sky. The inferred post-trial $p$-value is 18.8\%. Right: The time-independent event weights, evaluated for the IC86 V-VII data in the Northern hemisphere, at a source direction $\vec{x}_s$ defined by the hottest spot (RA, Dec) = (77.65$^{\circ{}}$, 2.55$^{\circ{}}$). The best-fit Gaussian time PDF is shown in black (dashed), with mean $\hat{T}_0$ and sigma $\hat{\sigma}_0$.}
\label{fig:unblinding_result_north_2015_2017}
\end{center}
\end{figure}

\begin{figure}[!ht]
\begin{center}
\begin{minipage}{0.39\linewidth}

\includegraphics[width=1\textwidth]{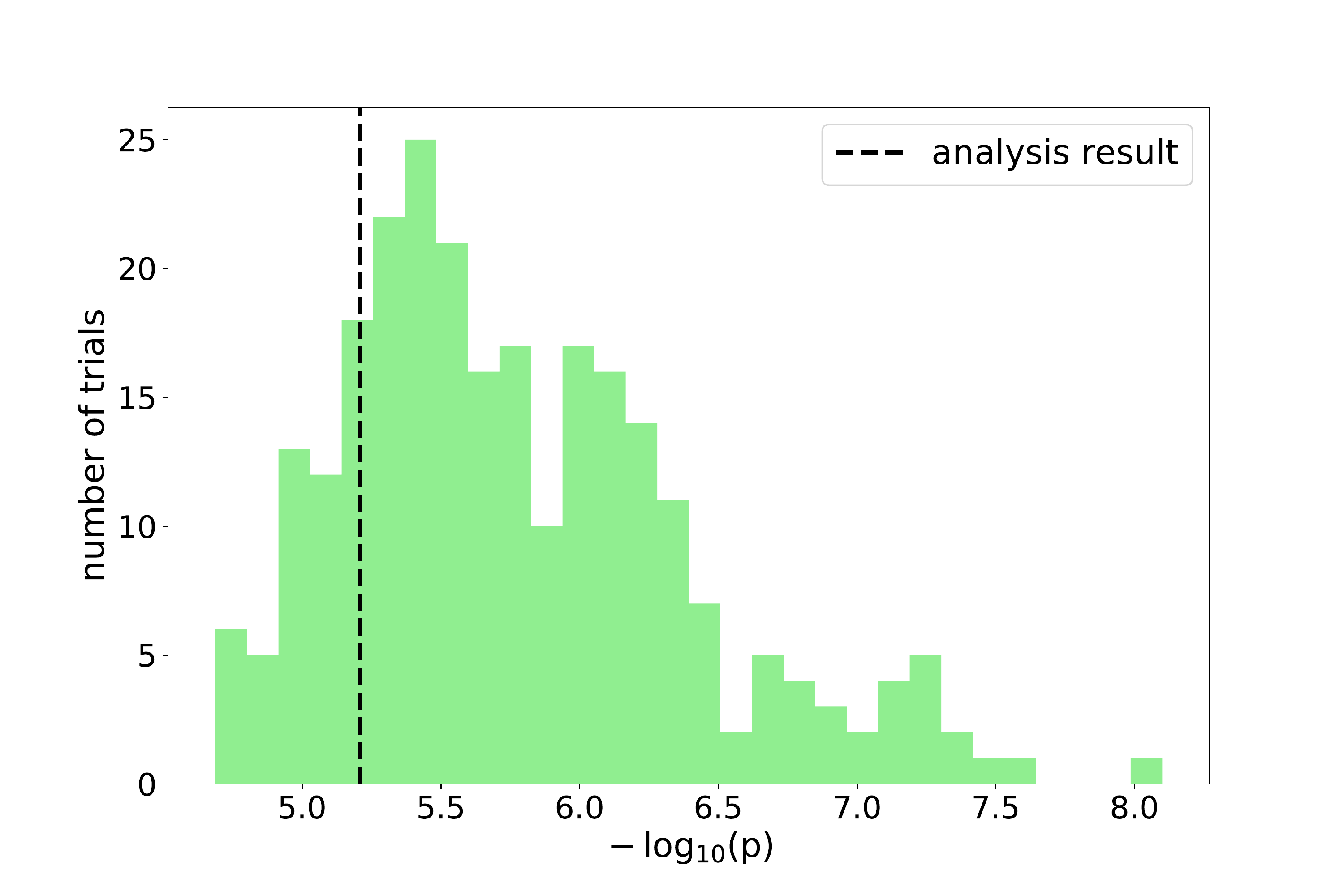}\\ 
\end{minipage}
\hfill
\begin{minipage}{0.59\linewidth}

\includegraphics[width=1.1\textwidth]{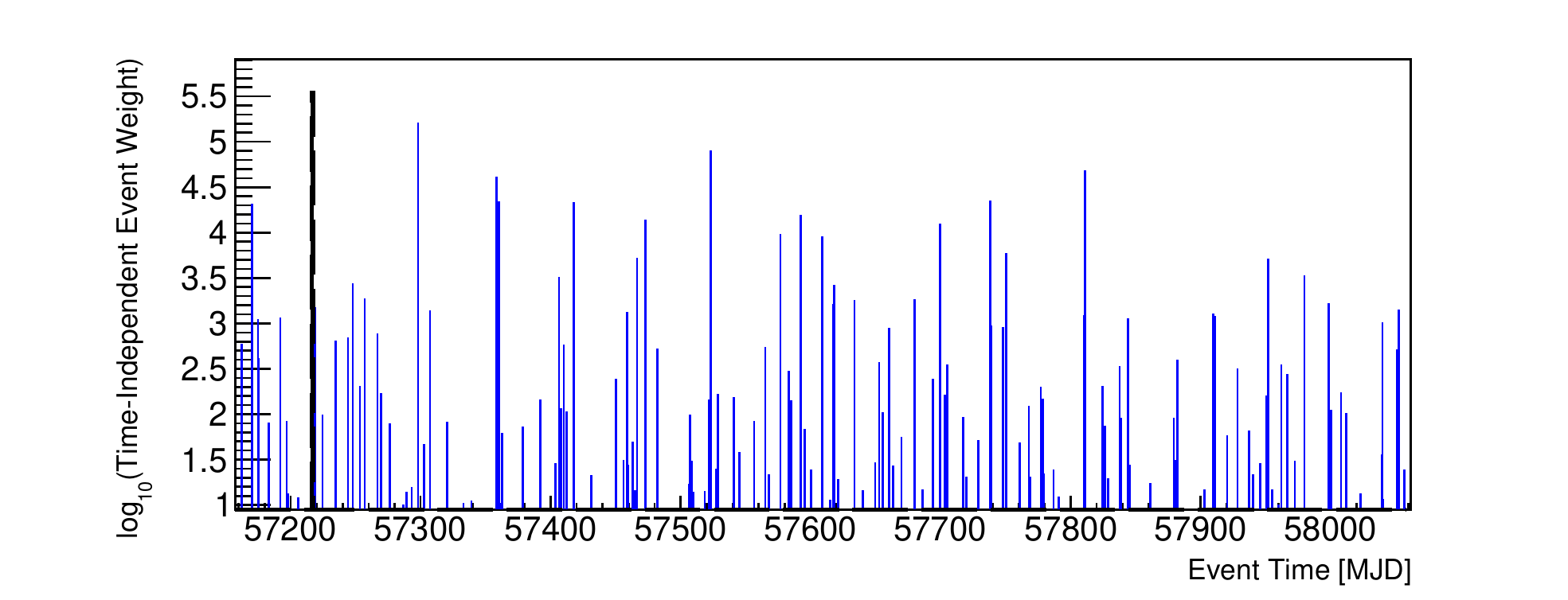}\\ 
\end{minipage}
\caption{Southern hemisphere: Left: expected background $p$-value distribution obtained from scrambled data in green compared to the measured most significant pre-trial $p$-value (shown as the black vertical dashed line) in the Southern sky. The inferred post-trial $p$-value is 81.5\%. Right: The time-independent event weights, evaluated for the IC86 V-VII data in the Southern hemisphere, at a source direction $\vec{x}_s$ defined by the hottest spot (RA, Dec) = (9.85$^{\circ{}}$, -31.05$^{\circ{}}$). The best-fit Gaussian time PDF is shown in black (dashed), with mean $\hat{T}_0$ and sigma $\hat{\sigma}_0$.
}
\label{fig:unblinding_result_south_2015_2017}
\end{center}
\end{figure}

The right panel of Fig.~\ref{fig:unblinding_result_north_2015_2017} (\ref{fig:unblinding_result_south_2015_2017}) shows the time-independent weights in blue, at a source direction $\vec{x}_s$ defined by the hottest spot in the Northern (Southern) hemisphere and the associated best-fit Gaussian in black. The corresponding histograms of background trials and the observed $p$-value (vertical black dashed line) are displayed in Fig.~\ref{fig:unblinding_result_north_2015_2017} (\ref{fig:unblinding_result_south_2015_2017}) for the Northern (Southern) hemisphere.

In Fig.~\ref{fig:unblinding_result_north_2015_2017}, we can see that the best-fit Gaussian encompasses almost the entire analysis period. Given that no finer structure is found, the fit tries to maximize the test statistics by including as much background as possible, hence the largest Gaussian width is preferred.

No significant excess is found in the Southern hemisphere (Fig.~\ref{fig:unblinding_result_south_2015_2017}) with a post-trial $p$-value of more than 80\%. 

Tab. \ref{tab:all_sky_scans_results} summarizes the results for each data period and each hemisphere, listing the coordinates of the hottest spot, the best-fit parameters and the pre-trial and post-trial $p$-values. In both cases, the smallest $p$-values were observed in the Northern hemisphere, where a longer duration flare, with a harder emission spectrum, was found. None of those results is significant.

\begin{deluxetable*}{ccccccccc}[ht!]
\tablecaption{Summary of all-sky scan results. \label{tab:all_sky_scans_results}}
\tablecolumns{6}
\tablewidth{0pt}
\tablehead{
\colhead{Data set} &
\colhead{RA (J2000)} &
\colhead{Dec (J2000)} & \colhead{Number of signal events} & \colhead{Spectral index} & \colhead{Gaussian mean} & \colhead{Gaussian width} & \colhead{p-value} & \colhead{p-value} \\
 & \colhead{($^\circ$)} & \colhead{($^\circ$)} & \colhead{$\hat{n}_s$} & \colhead{$\hat{\gamma}$} & \colhead{$\hat{T}_0$ (MJD)} & \colhead{$\hat{\sigma}_T$ (days)} & \colhead{pre-trial} & \colhead{post-trial}
}
\startdata
IC86 II-IV & & & & & & & \\
South	& 89.45  & -35.95  & 7.48 & 2.85 & 56808.46 & 4.013 & $1.38\times 10^{-6}$ & 24.2\% \\
North	& 170.35 &  27.95 & 11.79  & 2.11 & 56167.27 & 40.01 & $1.14\times 10^{-6}$ & 17.7\% \\
\hline
IC86 V-VII & & & & & & & \\
South & 9.85 & -31.05 & 2.99 & 3.55 & 57216.78 & 0.021 & $ 6.19 \times 10^{-6}$ & 81.5\% \\
North	& 77.65 & 2.55 & 25.27 & 2.25 & 57573.85 & 189.6 & $1.3 \times 10^{-6}$ & 18.8\%
\enddata
\end{deluxetable*}

\subsubsection{Results of the scans at the location of TXS 0506+056}

\citet{IceCube:2018cha} performed a dedicated time-dependent analysis at the position of the blazar TXS 0506+056, (RA, Dec,) = ($77.4^\circ{}$, $+5.7^\circ{}$), motivated by the detection of IceCube-170922A, and found a $3.5\sigma$ excess (post-trial $p$-value = $3\times 10^{-5}$), corresponding to $13\pm 5$ signal events in the time period between 2012 and 2015.
The excess doesn't appear significant in the present analysis, even though the data and the analysis method are the same. This is explained by the fact that this work scans the entire sky while \citet{IceCube:2018cha} is solely inspecting TXS 0506+056 direction. Since a grid of $0.1^\circ{} \times 0.1^\circ{}$ over the whole sky is analyzed, the final $p$-values are penalized by large trial factors, which reduce their significance. It is interesting to note that the second most significant hot spot (pre-trial $p$-value = $9.2\times 10^{-6}$) of the 2012-2015 analysis is found at (RA, Dec,) = ($77.55^\circ{}$, $+5.65^\circ{}$), which is consistent with the location of TXS 0506+056.
Moreover, applying the analysis at the exact TXS 0506+056 location as a further check, this work finds a similar excess of events ($\hat{n}_s = 13.55$) as reported by \citet{IceCube:2018cha} in the time period of 2012 to 2015, with a pre-trial $p$-value of $1.65 \times 10^{-5}$ and a pre-trial $p$-value of $3.2 \times 10^{-2}$ in the time period of 2015 to 2017, consistent with the pre-trial $p$-values found by \citet{IceCube:2018cha}.

\subsection{3C 279 flare analysis}

\begin{figure}[!ht]
\begin{center}
\begin{minipage}{0.39\linewidth}
\includegraphics[width=1.1\textwidth]{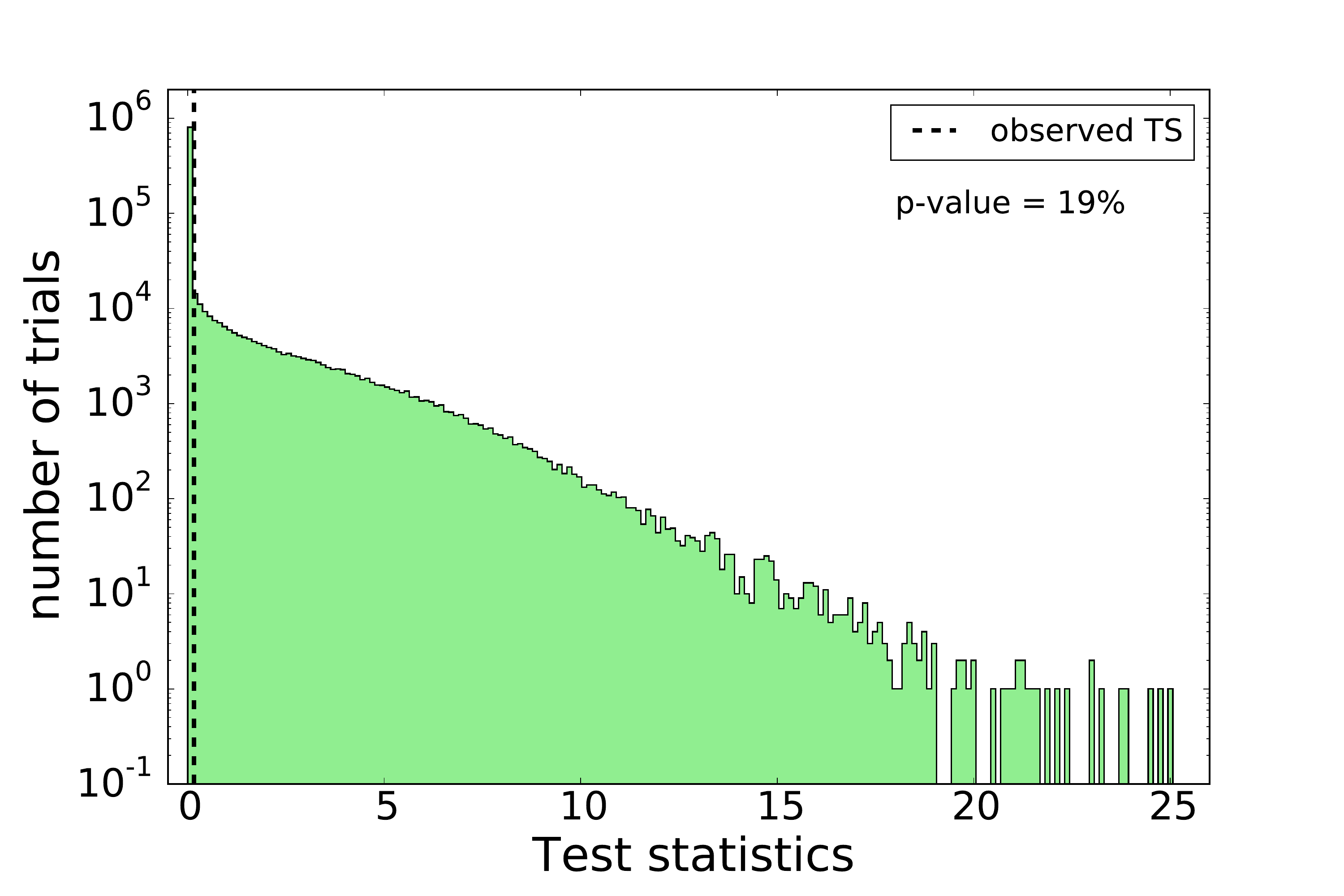}
\end{minipage}
\hfill
\begin{minipage}{0.59\linewidth}
\includegraphics[width=1\textwidth]{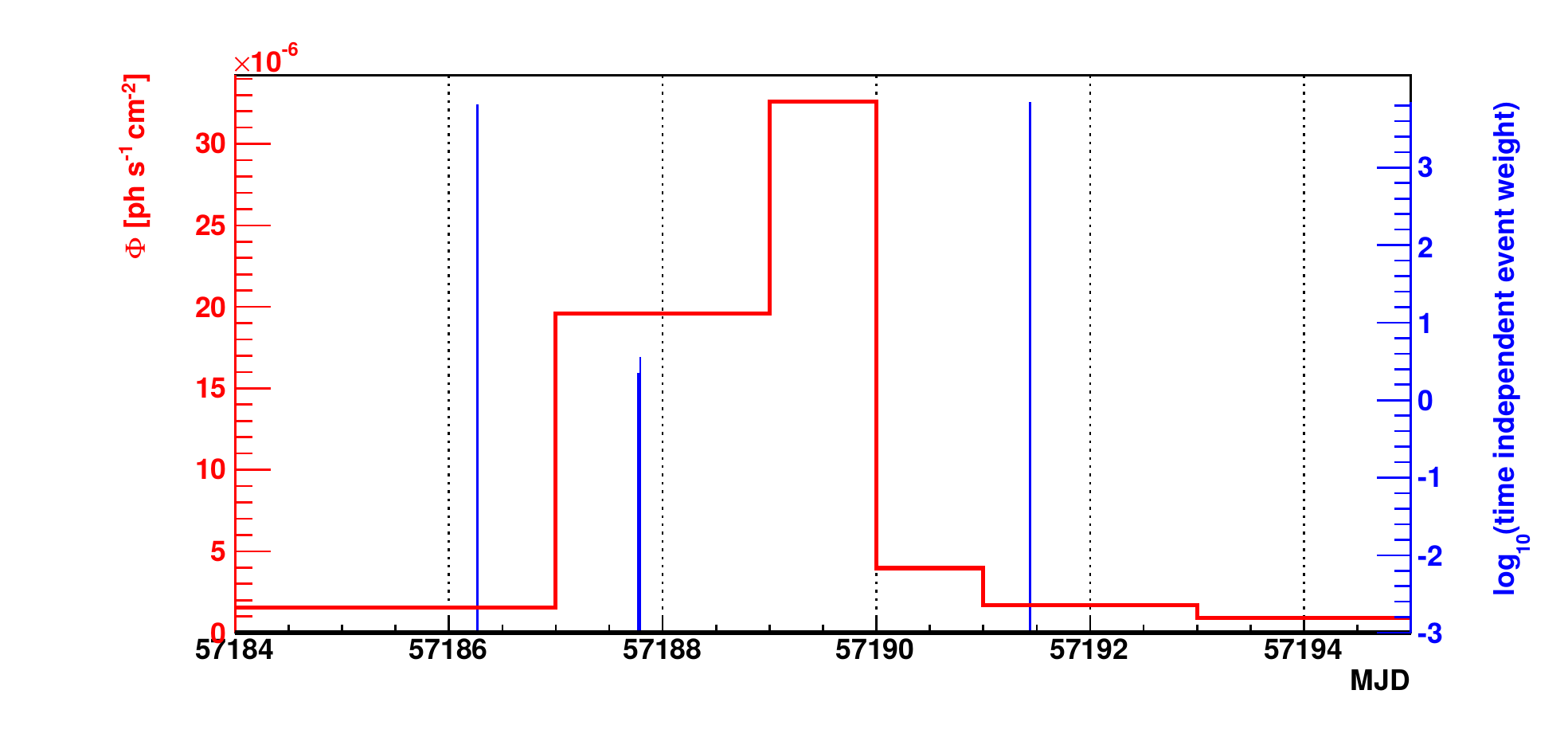}
\end{minipage}
\caption{Result of the 11-day (MJD 57184 - 57195 ) analysis. Left: test statistics for $10^6$ background trial scrambled data in green. The black dashed line shows the \textit{TS} value resulting from the analysis. Right: The red line is the denoised light curve of 3C 279. The blue vertical lines represent IceCube events selected by the analysis and their size is given by the product of the signal over the background spatial and energy PDFs. The threshold is not represented since it was fitted to zero. 
No correlation is found between the flare of 3C 279 and neutrinos detected by IceCube. 
}
\label{fig:unblinding_result_3C279}
\end{center}
\end{figure}

This analysis looks for high-energy neutrino candidate events clustered in space around 3C 279, which is located in the Southern sky at (RA, Dec)=($194.047^\circ{}$, $-5.789^\circ{}$), and in the time period defined by the blazar gamma-ray light curve. The result of the analysis is shown in Fig.~\ref{fig:unblinding_result_3C279}. The left panel displays the background \textit{TS} values found by running the analysis on $10^6$ scrambled data sets. The result of the analysis performed on real data is marked with the black vertical dashed line. A final $p$-value of 19\% is obtained, which is compatible with background. 

The right panel of Fig.~\ref{fig:unblinding_result_3C279} shows the time-independent neutrino event weights as defined by Eq.~\ref{eq:weights}. The best-fit value of the threshold is $\rm 2.44 \times 10^{-15}~photons~cm^{-2}~s^{-1}$, which is consistent with 0. This means that the time PDF is above this value and includes all 11 days. Tab.~\ref{tab:ana_results} summarizes the best-fit values of the parameters. The best-fit number of signal events is $\hat{n}_s=0.48$, the spectral index $\hat{\gamma}$ is 2.45 and the lag was fit to its maximum value, i.e. $0.5$ day. 

\begin{deluxetable*}{ccCccc}[ht!]
\tablecaption{Results of the 3C 279 flare analysis \label{tab:ana_results}}
\tablecolumns{6}
\tablewidth{0pt}
\tablehead{
\colhead{Number of signal events} &
\colhead{Spectral index} &
\colhead{Threshold} & \colhead{Lag } & \colhead{Test statistic} & \colhead{$p$-value} \\
\colhead{$\hat{n}_s$} & \colhead{$\hat{\gamma}$} &
\colhead{(photons~cm$^{-2}$~s$^{-1}$)} & (days) & &
}
\startdata
0.48 & 2.45 & 0 & 0.5 & 0.16 & 19\% \\
\enddata
\end{deluxetable*}

Four events are visible in the right plot of Fig.~\ref{fig:unblinding_result_3C279}, but a total of nine contributed to the likelihood. However their time-independent weight is so small that they are not represented in the plot. One may notice that two events around 57188~MJD are quite close in time. More details about the nine events can be found in Tab.~\ref{tab:ana_results_events}, which shows the time, coordinates, weights and visible reconstructed muon energy of each detected event in the 11-day analysis. 
\begin{deluxetable*}{cCCccc}[h!]
\tablecaption{IceCube events contributing to the likelihood fit, at the position of 3C 279, in the time period MJD 57184 - 57195. \label{tab:ana_results_events}}
\tablecolumns{6}
\tablewidth{0pt}
\tablehead{
\colhead{Event time } &
\colhead{$w_i$} &
\colhead{RA (J2000)} & \colhead{Dec (J2000)} & \colhead{Angular error (1$\sigma$)} & \colhead{$\rm \log_{10}({E_{\mu}}(GeV))$} \\
\colhead{(MJD)} & \colhead{($\rm \log_{10}{(spatialWeight \cdot energyWeight)}$)} &
\colhead{($^\circ{}$)} & {($^\circ{}$)} & ($^\circ{}$) &
}
\startdata
57185.41 & -6.13 & 194.91 & -3.35 &  0.37 &  2.81 \\
57185.72 & -2.98 & 190.00 & 1.56  &  1.59 &  2.75 \\
57186.27 & 3.82  & 193.95 & -4.99 &  0.46 &  3.26 \\
57187.77 & 0.35  & 187.34 & -4.69 &  1.91 &  3.07 \\
57187.79 & 0.55  & 195.35 & -3.89 &  0.57 &  3.10 \\
57189.15 & -3.30 & 187.78 & 12.07 &  3.73 &  2.87 \\
57189.50 & -3.50 & 197.67 & -1.82 &  0.94 &  2.87 \\
57191.44 & 3.85  & 194.95 & -6.02 &  0.60 &  4.01 \\
57194.61 & -3.81 & 195.93 & 2.15  &  1.44 &  2.89 \\
\enddata
\end{deluxetable*}

\begin{figure}[!ht]
\begin{center}
\includegraphics[width=.7\textwidth]{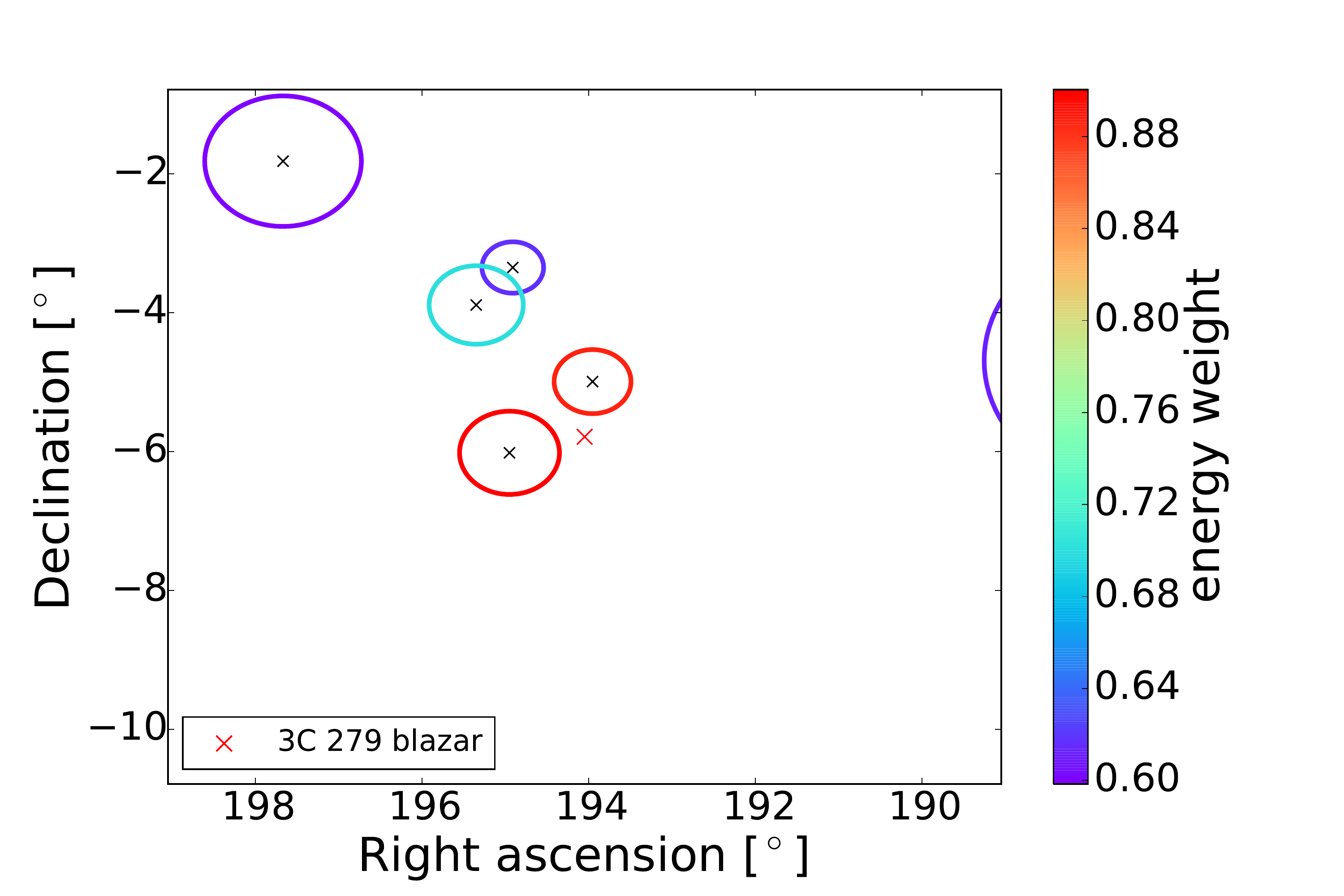}
\end{center}
\caption{Location in declination and right ascension of the IceCube events detected in the 11 days of the analysis. 3C 279 is represented by the red cross. The color scale shows the energy weight and the circles the angular uncertainty (i.e. paraboloid sigma, cf. Sec.~\ref{sec:ICdetectorDataSets}). Nine events are detected in total, but five have a weight close to zero, which probably means background}.
\label{fig:spatial_distri_events}
\end{figure}

The spatial distribution of the events found in $\pm 5^{\circ}$ around 3C 279 is shown in Fig.~\ref{fig:spatial_distri_events}. The position of 3C 279 is marked with a red cross. The color scale represents the energy weight while the circles shows the angular error associated with each event. The four events which are the closest to the source have the highest energy and a fairly good angular precision and are the ones seen in Fig~\ref{fig:unblinding_result_3C279}.

\subsection{Upper limit on a hadronic model} \label{subsec:3C279_limits}

The number of neutrinos expected to be detected at IceCube in association to the huge flare of 3C 279 is estimated by \citet{halzen2016high}, for a purely hadronic gamma-ray production mechanism. Considering the one day of highest emission of 3C 279 and taking into account IceCube's efficiency, the authors find that about 4 (2) events should be detected in time and space coincidence with the blazar flare for $pp$ ($p\gamma$) interactions. Since this analysis did not find any excess of neutrinos following the gamma-ray time profile from the direction of the blazar, we can constrain the emission model of~\citet{halzen2016high} by setting an upper limit on the number of neutrino events emitted by the source. 

\citet{halzen2016high} assume that the observed gamma rays in the MeV-TeV energy range are produced in a hadronic scenario, i.e. from protons colliding with radiation or gas surrounding the blazar. In this mechanism, charged and neutral pions are created and decay in turn, producing neutrinos and high-energy gamma rays. In order to derive the expected number of neutrinos that should be detected in IceCube, the neutrino flux has first to be related to the observed gamma-ray flux. Following estimates \citep{Halzen:2005pz} and using energy conservation, one can write:

\begin{equation}
\label{eq:enConservation}
    \int_{E_\gamma^{\text{min}}}^{E_\gamma^{\text{max}}} E_\gamma \frac{dN_\gamma}{dE_\gamma} dE_\gamma = K \int_{E_\gamma^{min}}^{E_\gamma^{max}} E_\nu \frac{dN_\nu}{dE_\nu} dE_\nu 
\end{equation}

where the gamma-ray spectrum is defined as 

\begin{equation}
\label{eq:gammaRaySpectrum}
    \frac{dN_\gamma}{dE_\gamma} = A_\gamma E_\gamma^{-\alpha}
\end{equation}
where 
\begin{equation}
A_\gamma = \frac{F_\gamma}{\int_{E_\gamma^{\text{min}}}^{E_\gamma^{\text{max}}} E_\gamma^{-\alpha} dE_\gamma}
\end{equation}

with $\alpha=2.1$ the measured spectral index, $\rm F_\gamma = 24.3\times10^{-6}~photons~cm^{-2}~s^{-1}$ the average daily flux during the one day of highest emission \citep{ackermann2016minute}, $E_\gamma^{\text{min}}=0.1$~GeV and $E_\gamma^{\text{max}}=300$~GeV.
The factor $K$ indicates the ratio between the number of gamma rays and the number of neutrinos. For proton interactions with radiation, $K_{p\gamma}=\frac{\text{BR}_{\gamma}}{\text{BR}_{\nu}}=\frac{5/8}{3/8}=1.7$ and for $pp$ interactions, $K_{pp}=\frac{\text{BR}_{\gamma}}{\text{BR}_{\nu}}=\frac{1/2}{1/2}=1$. $\rm BR_\gamma$ and $\rm BR_\nu$ are the branching ratio of gamma rays and neutrinos, respectively. They are calculated in the same way as in \citet{halzen2016high}, the only difference being that we also consider synchrotron radiation or Compton scattering from the positrons produced by charged pions, which produces additional gamma rays.

The neutrino spectrum is derived from Eq.~\ref{eq:enConservation}:
\begin{equation}
\label{eq:neutrinoFlux}
 \frac{dN_\nu}{dE_\nu} \approx A_\nu E_\nu^{-2} \approx \frac{A_\gamma E_{\gamma,min}^{-\alpha+2}}{(\alpha-2)K\ln(E_{\nu,\text{max}}/E_{\nu,\text{min}})} E_{\nu}^{-2}   
\end{equation}

Following the steps described in  \cite{halzen2016high} and integrating Eq.~\ref{eq:neutrinoFlux}, the expected number of muon neutrinos in IceCube is 

\begin{equation}
    N_{\nu_{\mu}+\bar{\nu}_{\mu}} = t \int_{E_\nu^{\text{min}}}^{E_\nu^{\text{max}}}  \frac{1}{3}\frac{dN_\nu}{dE_\nu} A_{\text{eff}}(E_\nu,\delta)dE
    \label{eq:numberEvents}
\end{equation}
where we choose $t=86400$~s, $1$~TeV$\leq E_{\nu}\leq $10~PeV and the declination band of 3C 279, $-6.289^\circ{} \leq \delta \leq -5.289^\circ{}$. The factor $\frac{1}{3}$ accounts for neutrino oscillations. $A_{\text{eff}}(E_\nu,\delta)$ is IceCube effective area at the declination of 3C 279 and is shown by the dashed green line in Fig.~\ref{fig:effective_area}.

In order to compare against the model of \citet{halzen2016high}, a 90\% C.L. upper limit is calculated with the 3C 279 flare analysis. In this calculation, the injected signal is defined by a point-source at the location of 3C 279 emitting a flux proportional to $E^{-2}$, having fixed a high threshold in order to effectively reduce the emission time to the same assumed in Ref.~\citet{halzen2016high}, corresponding to 1 day on 2015 June 16, when the highest emission was observed. The signal strength at which 90\% of pseudo-experiments had a larger \textit{TS} than the observed value corresponds to a mean of 2.18 signal events, which is slightly larger than the number of events needed to reach the sensitive region. This number of events translates to a time-integrated flux of $ E^{2}\frac{1}{A_{\text{eff}}} \frac{dN}{dE} \rm = 4.24 \cdot 10^{-2}~GeV/cm^{2}$.

\begin{figure}[!ht]
\begin{center}
\begin{minipage}{0.4\linewidth}
\centering \includegraphics[scale=0.3]{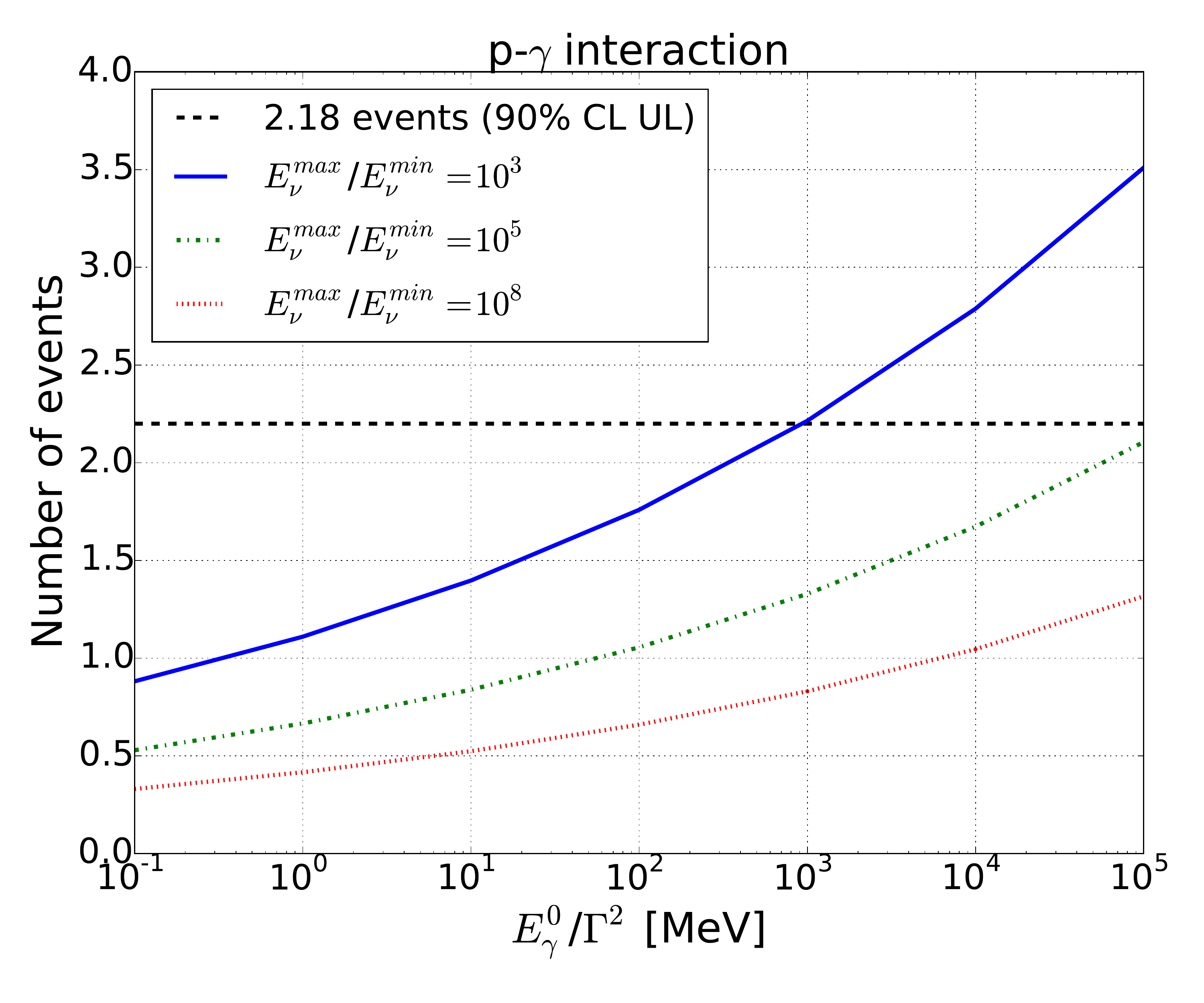}
\end{minipage}
\hfill
\begin{minipage}{0.4\linewidth}
\centering \includegraphics[scale=0.3]{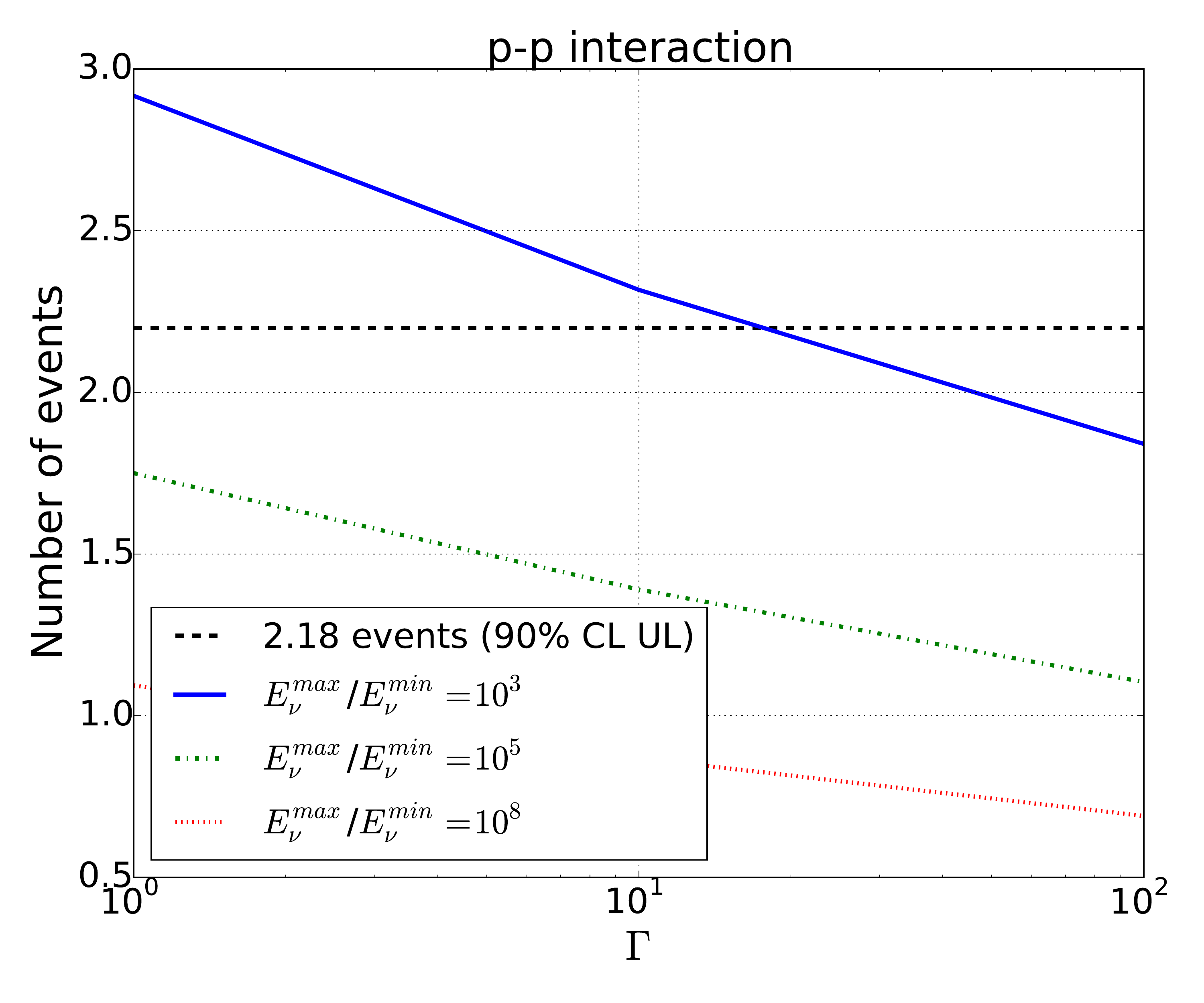}
\end{minipage}
\caption{Left: Number of expected muon neutrino tracks in IceCube for $p\gamma$ interactions vs. $E_{\gamma}^{o}/\Gamma^2$ , with $E_{\gamma}^{o}$ the energy of the target photons as derived from \citet{halzen2016high}. The different colors represent different energy ranges where neutrinos are produced. The black dashed line represents the 90\% CL upper limit corresponding to the one-day block on June 16 derived from the analysis. The parameter space above the dashed line is excluded.
Right: Number of expected muon neutrino tracks in IceCube vs. jet Lorentz factor $\Gamma$ in the case of $pp$ interactions, as derived in~\citet{halzen2016high}. The different colors represent different neutrino production energy ranges. The black dashed line represents the 90\% CL upper limit corresponding to the one-day block on June 16 derived from the analysis. The parameter space above the dashed line is excluded.}
\label{fig:model_constraints}
\end{center}
\end{figure}

\begin{figure}[!ht]
\begin{center}
\begin{minipage}{0.49\linewidth}
\centering \includegraphics[scale=0.3]{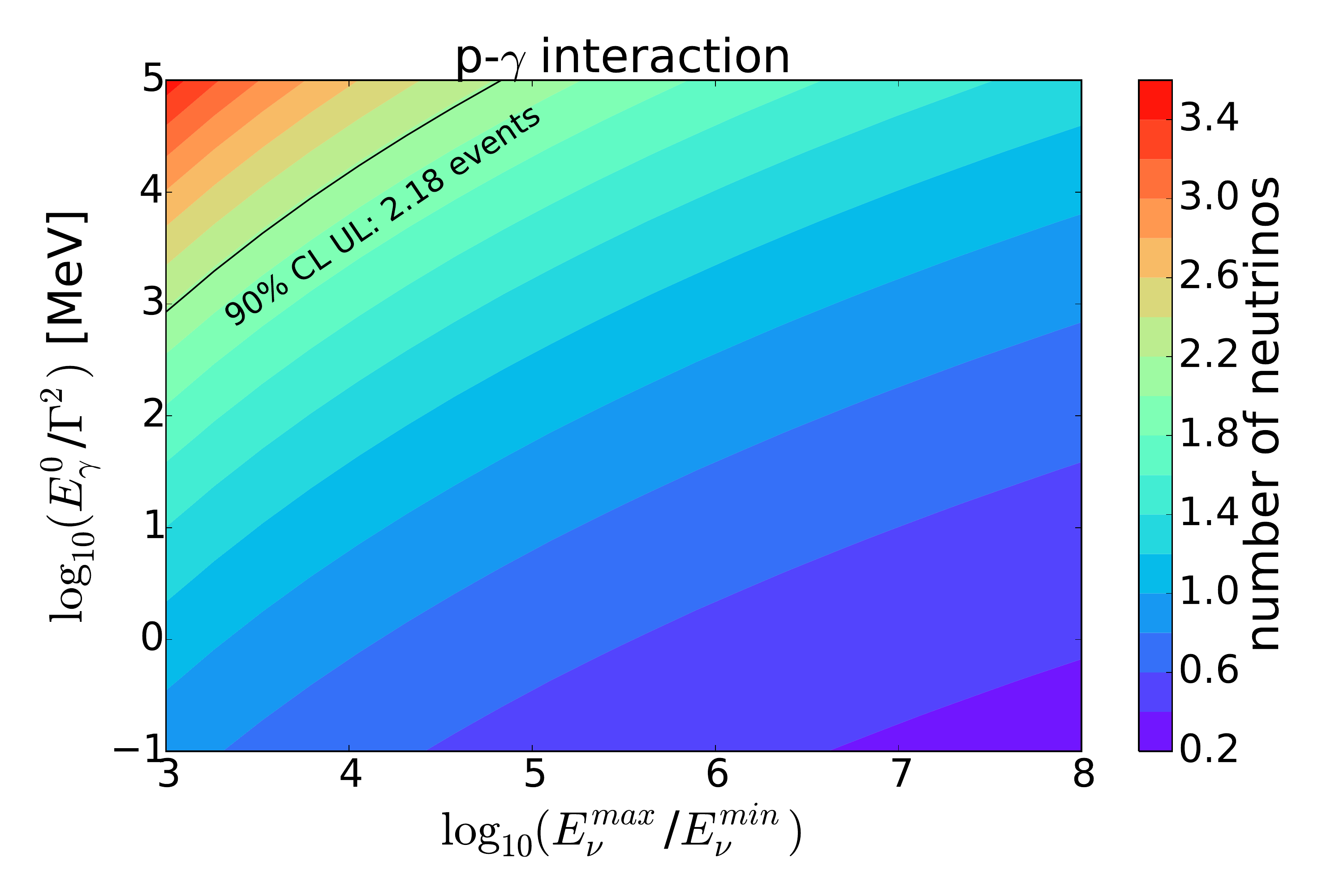}
\end{minipage}
\hfill
\begin{minipage}{0.49\linewidth}
\centering \includegraphics[scale=0.3]{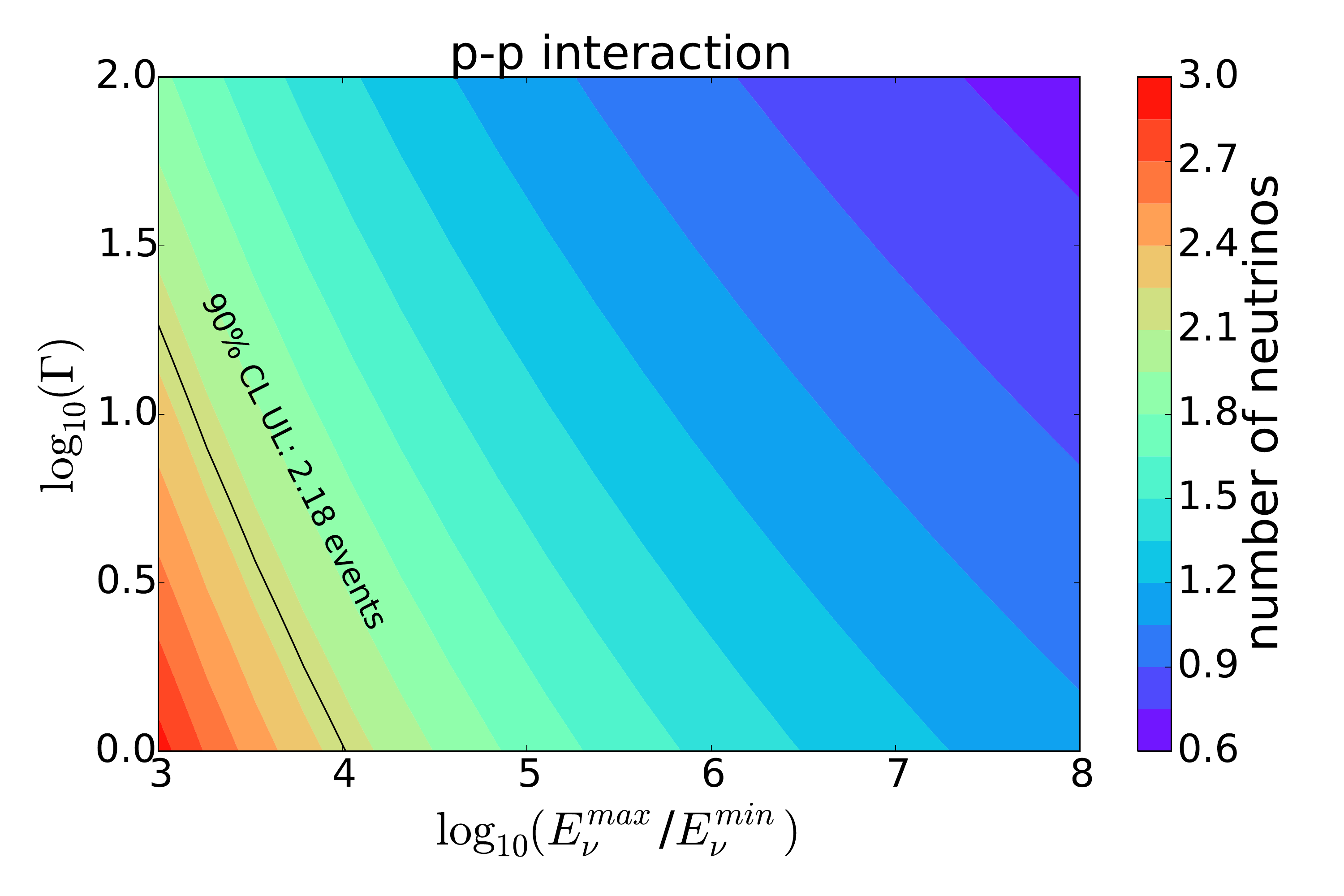}
\end{minipage}
\caption{ Contour plots showing the expected number of muon neutrino tracks in IceCube for $p\gamma$ (left) and for $pp$ (right) interactions. The black line represents the 90\% CL upper limit  corresponding to the one-day block on June 16 derived from the analysis. The parameter space on the left of the black line is excluded.}
\label{fig:model_constraints_2d}
\end{center}
\end{figure}

The expected number of muon neutrino tracks in IceCube, $N_{\nu_{\mu}+\bar{\nu}_{\mu}}$, is calculated from Eq.~\ref{eq:numberEvents}. Following the approach of \citet{halzen2016high}, different neutrino energy ratios, $E_{\nu,\text{max}}/E_{\nu,\text{min}}$, which represent the energy interval over which proton interactions produce gamma rays from pions, are tested and shown in Fig.~\ref{fig:model_constraints} and Fig.~\ref{fig:model_constraints_2d}. The minimum neutrino energy, $E_{\nu,\text{min}}$ is directly derived from the minimum energy of the protons, which dependins on the value of the Lorentz factor of the jet of the blazar that is considered. The exact calculations are described by Eq. 4 to 7 in \citet{halzen2016high}. Once $E_{\nu,\text{min}}$ is set, $E_{\nu,\text{max}}$ is determined from the chosen neutrino energy ratio. 

Fig.~\ref{fig:model_constraints}~shows the number of expected muon neutrino tracks in IceCube as a function of the Lorentz factor of the jet of the blazar $\Gamma$ in the case of $pp$ interactions (right) and as a function of $E_{\gamma}^{o}/\Gamma^2$ , with $E_{\gamma}^{o}$ the energy of the target photons, in the case of $p\gamma$ interactions (left). The blue line corresponds to $E_{\nu,\text{max}}/E_{\nu,\text{min}}=10^3$, the green line to $E_{\nu,\text{max}}/E_{\nu,\text{min}}=10^5$ and the red line to $E_{\nu,\text{max}}/E_{\nu,\text{min}}=10^8$. The black dashed line represents the 90\% CL upper limit corresponding to the one-day block on June 16 derived from this analysis. Model parameters which lie above the dashed line are excluded. 

Only a small fraction of the parameter space can be excluded by this analysis, assuming the smallest ratio $E_{\nu,\text{max}}/E_{\nu,\text{min}}$, which represents the energy interval over which proton interactions produce gamma rays from pions. In the case of $p\gamma$ interactions large $E_{\gamma}^{o}/\Gamma^2$ ($>10^3$) can be excluded, while for the $pp$ scenario small $\Gamma$ ($<20$) are disfavored. 

Fig.~\ref{fig:model_constraints_2d} gives the same information as Fig.~\ref{fig:model_constraints} in two dimensions. The colored regions correspond to the expected number of events. The parameter spaces on the left of the black lines are excluded.

\newpage
\section{Summary}

In this paper, we reported a time-dependent all-sky scan on IceCube data from 2012 to 2017. The analysis searches each direction in the sky for an excess of neutrino events in space and time above the expected background. No assumptions on the potential sources of neutrinos are made. The most significant result corresponds to a $p$-value of 17.7\%, which is compatible with background. It is obtained at the beginning of the analyzed period, the best-fit Gaussian time structure having its mean on 2012 August 28 and a width of 40 days, and lies in the Northern hemisphere at the position (RA, Dec) = ($170.4^\circ, 28.0^\circ$). 

Within the analyzed period, the blazar 3C 279, already well-known for its high variability, went through a brief but extremely intense gamma-ray flare which was recorded by the \textit{Fermi}-LAT experiment. In order to analyze this interesting event, a dedicated time-dependent search was performed at the location of the blazar and during 11 days around the peak of the flare, which happened on 2015 June 16. Assuming that the neutrino emission would follow the intensity of the detected gamma rays, we defined the signal time PDF directly from the gamma-ray light curve of 3C 279, obtained from \textit{Fermi}-LAT. The analysis resulted in a $p$-value of 19\%, which is compatible with background.

The 90\% C.L. upper limit for the day of most intense gamma-ray activity of the blazar was calculated and used to constrain the neutrino emission model presented in \citet{halzen2016high}. This model, which is purely hadronic, estimates the number of neutrinos that should be detected by IceCube as function of the Lorentz boost of the jet of the blazar $\Gamma$ and of the energy of the target photons $E_{\gamma}^{o}$. Different neutrino energy intervals over which proton interactions would produce gamma rays from pions were calculated and the results were expressed as a function of the neutrino energy ratio, $E_{\nu,\text{max}}/E_{\nu,\text{min}}$. Assuming the smallest ratio of neutrino energies leads to the most stringent limits, where large $E_{\gamma}^{o}/\Gamma^2$ ($>10^5$) are excluded in case of p$\gamma$ interactions, while small $\Gamma$ ($<20$) are disfavored for the $pp$ scenario. 

\section*{Acknowledgements}
The IceCube collaboration acknowledges the significant contributions to this manuscript from St\'ephanie Bron and Teresa Montaruli. The authors gratefully acknowledge the support from the following agencies and institutions:
USA {\textendash} U.S. National Science Foundation-Office of Polar Programs,
U.S. National Science Foundation-Physics Division,
Wisconsin Alumni Research Foundation,
Center for High Throughput Computing (CHTC) at the University of Wisconsin{\textendash}Madison,
Open Science Grid (OSG),
Extreme Science and Engineering Discovery Environment (XSEDE),
Frontera computing project at the Texas Advanced Computing Center,
U.S. Department of Energy-National Energy Research Scientific Computing Center,
Particle astrophysics research computing center at the University of Maryland,
Institute for Cyber-Enabled Research at Michigan State University,
and Astroparticle physics computational facility at Marquette University;
Belgium {\textendash} Funds for Scientific Research (FRS-FNRS and FWO),
FWO Odysseus and Big Science programmes,
and Belgian Federal Science Policy Office (Belspo);
Germany {\textendash} Bundesministerium f{\"u}r Bildung und Forschung (BMBF),
Deutsche Forschungsgemeinschaft (DFG),
Helmholtz Alliance for Astroparticle Physics (HAP),
Initiative and Networking Fund of the Helmholtz Association,
Deutsches Elektronen Synchrotron (DESY),
and High Performance Computing cluster of the RWTH Aachen;
Sweden {\textendash} Swedish Research Council,
Swedish Polar Research Secretariat,
Swedish National Infrastructure for Computing (SNIC),
and Knut and Alice Wallenberg Foundation;
Australia {\textendash} Australian Research Council;
Canada {\textendash} Natural Sciences and Engineering Research Council of Canada,
Calcul Qu{\'e}bec, Compute Ontario, Canada Foundation for Innovation, WestGrid, and Compute Canada;
Denmark {\textendash} Villum Fonden and Carlsberg Foundation;
New Zealand {\textendash} Marsden Fund;
Japan {\textendash} Japan Society for Promotion of Science (JSPS)
and Institute for Global Prominent Research (IGPR) of Chiba University;
Korea {\textendash} National Research Foundation of Korea (NRF);
Switzerland {\textendash} Swiss National Science Foundation (SNSF);
United Kingdom {\textendash} Department of Physics, University of Oxford.

\newpage

\bibliography{biblio}


\end{document}